\documentclass[fleqn,usenatbib]{mnras}
\usepackage{amsmath}  
\usepackage{amssymb}  

\usepackage{newtxtext,newtxmath}

\usepackage[T1]{fontenc}
\usepackage{ae,aecompl}

\usepackage{graphicx} 
\usepackage{cprotect}

\usepackage{color}

\def\mr#1{\mathrm{#1}}

\title[Substructures in the core of Abell~2319]{Substructures in the core of Abell~2319}

\author[Y. Ichinohe et al.]{
Y. Ichinohe,$^{1}$\thanks{E-mail: ichinohe@rikkyo.ac.jp}
A. Simionescu,$^{2,3,4}$
N. Werner,$^{5}$
M. Markevitch,$^{6}$
and Q. H. S. Wang$^{7}$
\\
$^{1}$Department of Physics, Rikkyo University, 3-34-1 Nishi-Ikebukuro, Toshima-ku, Tokyo 171-8501, Japan\\
$^{2}$SRON Netherlands Institute for Space Research, Sorbonnelaan 2, NL-3584 CA Utrecht, the Netherlands\\
$^{3}$Leiden Observatory, Leiden University, PO Box 9513, 2300 RA Leiden, The Netherlands\\
$^{4}$Kavli Institute for the Physics and Mathematics of the Universe (WPI), The University of Tokyo, Kashiwa, Chiba 277-8583, Japan\\
$^{5}$Department of Theoretical Physics and Astrophysics, Faculty of Science, Masaryk University, Kotlarsk\'{a} 2, Brno CZ-611 37, Czech Republic\\
$^{6}$NASA/Goddard Space Flight Center, Greenbelt, MD 20771, USA\\
$^{7}$Department of Physics and Astronomy, University of Utah, 115 South 1400 East, Salt Lake City, UT 84112, USA
}

\date{\today}

\pubyear{2020}

\begin{document}
\label{firstpage}
\pagerange{\pageref{firstpage}--\pageref{lastpage}}
\maketitle

\begin{abstract}
We analysed the deep archival {\it Chandra}\/ observations of the high-temperature galaxy cluster Abell~2319 to investigate the prominent cold front in its core. The main sharp arc of the front shows wiggles, or variations of the radius of the density jump along the arc. At the southern end of the arc is a feature that resembles a Kelvin-Helmholtz (KH) eddy, beyond which the sharp front dissolves. These features suggest that KH instabilities develop at the front. Under this assumption, we can place an upper limit on the ICM viscosity that is several times below the isotropic Spitzer value. Other features include a split of the cold front at its northern edge, which may be another KH eddy. There is a small pocket of hot, less-dense gas inside the cold front, which may indicate a `hole' in the front's magnetic insulation layer that lets the heat from the outer gas to penetrate inside the front. Finally, a large concave brightness feature southwest of the cluster core can be caused by the gasdynamic instabilities. We speculate that it can also be an inner boundary of a giant AGN bubble, similar to that in Ophiuchus. If the latter interpretation is supported by better radio data, this could be a remnant of another extremely powerful AGN outburst.
\end{abstract}




\section{Introduction}
Since its launch in 1999, {\it Chandra} has revealed a large number of fine X-ray morphological substructures in the intracluster medium (ICM), the most dominant baryonic component in clusters of galaxies. ICM substructure is a consequence of various cluster activities such as interactions with the central active galactic nucleus (AGN), mergers with other clusters or groups, and motions of its member galaxies. In the formation of these substructures, the ICM follows the physical laws, and thus, ICM substructures reflect the underlying physical properties of the ICM such as magnetic field strengths, viscosity and heat conductivity, little of which is well understood so far.

Among the substructures discovered by {\it Chandra} are cold fronts, the interface between two gas phases in pressure equilibrium: cooler and denser gas, and the hot and thin ambient medium \citep[see][for a review]{markevitch07}. The first cold fronts were found in Abell~2142 \citep{markevitch00} and in Abell~3667 \citep{vikhlinin01a,vikhlinin01b}. Since then, cold fronts have been observed in a variety of locations and environmental setups \citep[e.g.,][]{fabian06,machacek06,owers09b,ghizzardi10} and it has been revealed that they are actually more common than shock fronts.

One of the important properties of cold fronts is their remarkable sharpness. The thickness of the interface is not resolved even by {\it Chandra}'s angular resolution \citep{vikhlinin01a,vikhlinin01b}. This indicates that diffusive processes are suppressed at the interface because of, e.g., the small gyroradii of the electrons resulting from the draping magnetic fields.

Initially, cold fronts were regarded as single distinct substructures. However, recently, substructures of such substructures, i.e., substructures associated with cold fronts have been attracting attention as the tool to infer the underlying ICM microphysics. For example, \citet{werner16a} reported the presence of quasi-linear features underneath the cold front in the Virgo cluster. By comparison with a tailored numerical simulation, they suggested that these features are due to the amplification of magnetic fields by gas sloshing. Similar narrow structures of surface brightness were also found in several other systems; possible plasma depletion layers seen in projection ({\it X-ray channels}) were found in Abell~520 and Abell~2142 \citep[][]{wang16,wang18}; \citet{ichinohe19a} found alternating bright and faint regions ({\it feathers}) in the Perseus cluster and estimated the amplified magnetic field strength at $\sim$30\,$\mu$G. Several simulation studies have also been performed for magnetized gas sloshing \citep{zuhone11,zuhone15}.

Besides the features {\it inside} cold fronts, features {\it on} the cold front have also been extensively studied. As gas shear is inevitable at the cold front interface, fluid instabilities can develop under certain conditions. \citet{su17a} found a multiple-edge profile of the merger cold front in the NGC~1404 galaxy, which is likely attributed to Kelvin-Helmholtz instabilities (KHI) seen in projection \citep{roediger13a}. Similar multiple-edge surface brightness profiles were also found in Abell~3667 \citep{ichinohe17} and in the Perseus cluster \citep{ichinohe19a}. At small scales, the existence of KHI whose size of less than kpc is indicated in the core of the Ophiuchus cluster \citep{werner16b}, while at large scales, \citet{walker17} suggested that a KHI roll with a size of $\sim$50\,kpc exists east of the core of the Perseus cluster.

Abell~2319 ($z\sim0.056$; 1\,arcsec $\sim$1.08\,kpc) is the fifth brightest cluster in the X-ray sky \citep{edge90}. The early {\it Chandra} observations made by \citet{ohara04} and \citet{govoni04} revealed its high temperature and the existence of the sharp cold front extending for $\sim$300\,kpc to the southeast of the brightness peak of the system. Both of these features as well as its giant radio halo \citep{harris78,feretti97,govoni01,farnsworth13} indicate a major merger activity ongoing in this system, and there has been a wide range of studies in this context \citep{markevitch96,million09,sugawara09,yan14,storm15}. Abell~2319 is the most significant Sunyaev-Zel'dovich effect \citep[SZ;][]{sunyaev72} detection in the {\it Planck} 2nd Sunyaev-Zeldovich Source Catalog \citep{planck16xxvii} and detailed studies using {\it XMM-Newton} aiming at probing the outskirts of the cluster have been performed \citep{eckert14,ghirardini18,hurier19}.

There have been several studies in terms of the cold front itself; \citet{ghizzardi10} reconfirmed the existence of the front using {\it XMM-Newton} data; \citet{zuhone13a} used the {\it Chandra} image for visual comparison with numerical simulation images; \citet{walker16} used the {\it Chandra} image to demonstrate the Gaussian gradient magnitude filtering method. However, despite its prominence, there seem to have been no observational quantitative estimations of ICM microphysics performed using this cold front so far.

In this paper, we investigate in detail the currently available $\sim$90\,ks archival {\it Chandra} data of Abell~2319 to explore the phenomena associated with this remarkable cold front and the physical implications derived from it. The high temperature and moderate density environment around the cold front in Abell~2319 is advantageous compared to other brightest clusters in that the Coulomb mean free paths of the electrons are much longer in this system.

We adopted the abundance table of proto-solar metal from \citet{lodders09} for this paper. Unless otherwise noted, the error bars correspond to 68\% confidence level for one parameter. Throughout this paper, we assume the standard $\Lambda$CDM cosmological model with the parameters of $(\Omega_m,\Omega_\Lambda,H_0)=(0.3,0.7,70~\mr{km/s/Mpc})$.

\section{Observations, data reduction, and data analysis}\label{sec:data}
\begin{table}
 \centering
 \caption{Summary of the observations used in this paper. The net exposure time is after the data screening.}
 \label{tbl:data}
 \begin{tabular}{ccc}
  \hline
  Obs ID & Date & Net exposure time (ks)\\
  \hline
  3231 & 2002-03-15 &  14.4 \\
  15187 & 2014-02-11 &  75.0 \\
  \hline
 \end{tabular}
\end{table}

Abell~2319 was observed twice using the {\it Chandra} ACIS-I detectors (ObsIDs 3231 and 15187). We reprocessed the archival level 1 event lists produced by the {\it Chandra} pipeline in the standard manner\footnote{CIAO 4.12 Homepage, Data Preparation; http://cxc.harvard.edu/ciao/threads/data.html} using the {\small CIAO} software package (version 4.12) and the {\small CALDB} (version 4.9.2.1) to apply the appropriate gain maps and the latest calibration products. We removed flares from light curves using the \verb+deflare+ tool with the standard time binning method recommended in the {\small CIAO} official analysis guides. Blank-sky background files provided by the {\it Chandra} team were extracted using the \verb+blanksky+ tool and were processed in a similar manner. The net exposure times of each observation after screening are summarized in Table~\ref{tbl:data}. The resulting total net exposure time is $\sim$90\,ks.

\begin{figure}
  \centering
  \includegraphics[width=3.2in]{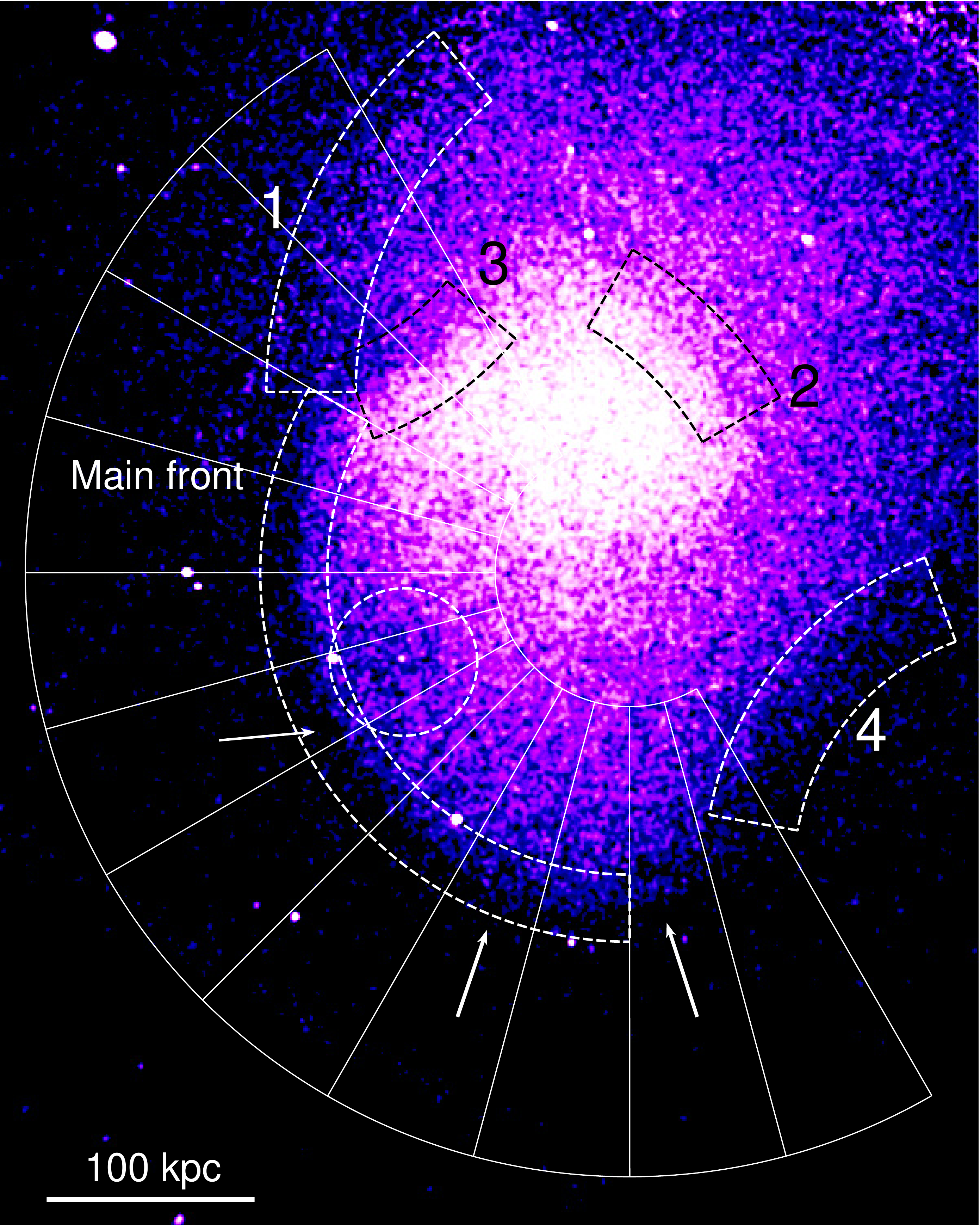}
 \caption[]{Exposure and vignetting corrected, background subtracted {\it Chandra} image (0.5--7.0\,keV) of Abell~2319, Gaussian smoothed with $\sigma=2$\,pixel. The main cold front is indicated by the white dashed partial annulus to southeast. The overlaid partial annuli shown with a solid line cover the azimuthal range from which the 15$^\circ$ surface brightness profiles are extracted (see also Section~\ref{sec:sbprof}). The radius of the cold front abruptly decreases/increases around the directions indicated by the left/middle white arrows. The right white arrow points the surface brightness `finger'. The numbered dashed partial annuli correspond to other interesting features. The white dashed circle denotes the surface-brightness depression. See Section~\ref{sec:morph} for details.}
 \label{img:flatimage}
\end{figure}

We created the count image and the exposure map, and the background image using the \verb+fluximage+ and \verb+blanksky_image+ tool, respectively. We combined all the images of both ObsIDs appropriately to create the exposure and vignetting corrected, background subtracted image (flat-fielded image). The resulting image is shown in Fig.~\ref{img:flatimage}. We identified point sources using the \verb+wavdetect+ tool with the scales of 1, 2, 4, 8, 16\,pixels and removed them in the subsequent analysis.

\begin{figure*}
 \begin{minipage}{0.33\hsize}
  \centering
  \includegraphics[width=2.2in]{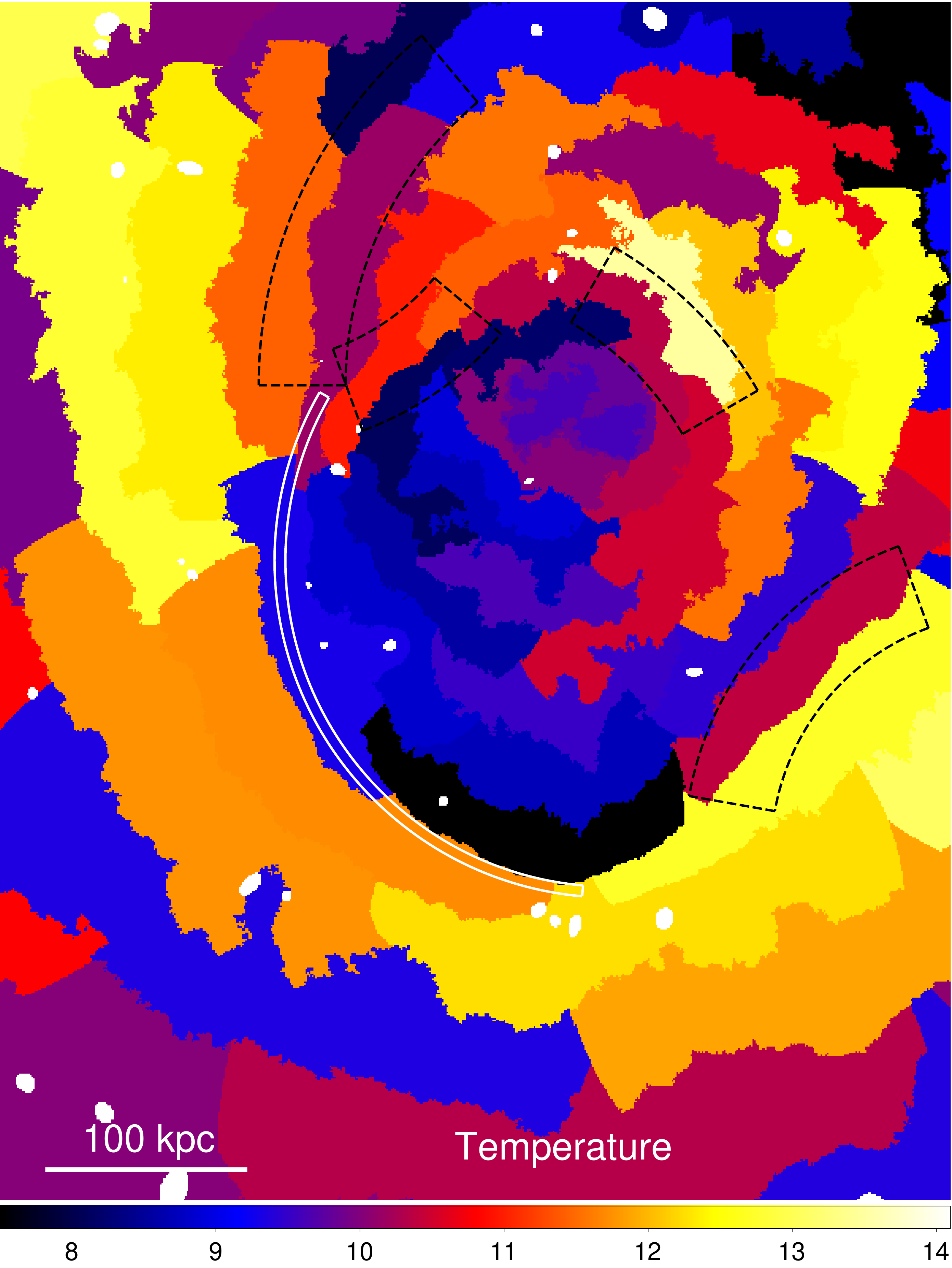}
 \end{minipage}
 \begin{minipage}{0.33\hsize}
  \centering
  \includegraphics[width=2.2in]{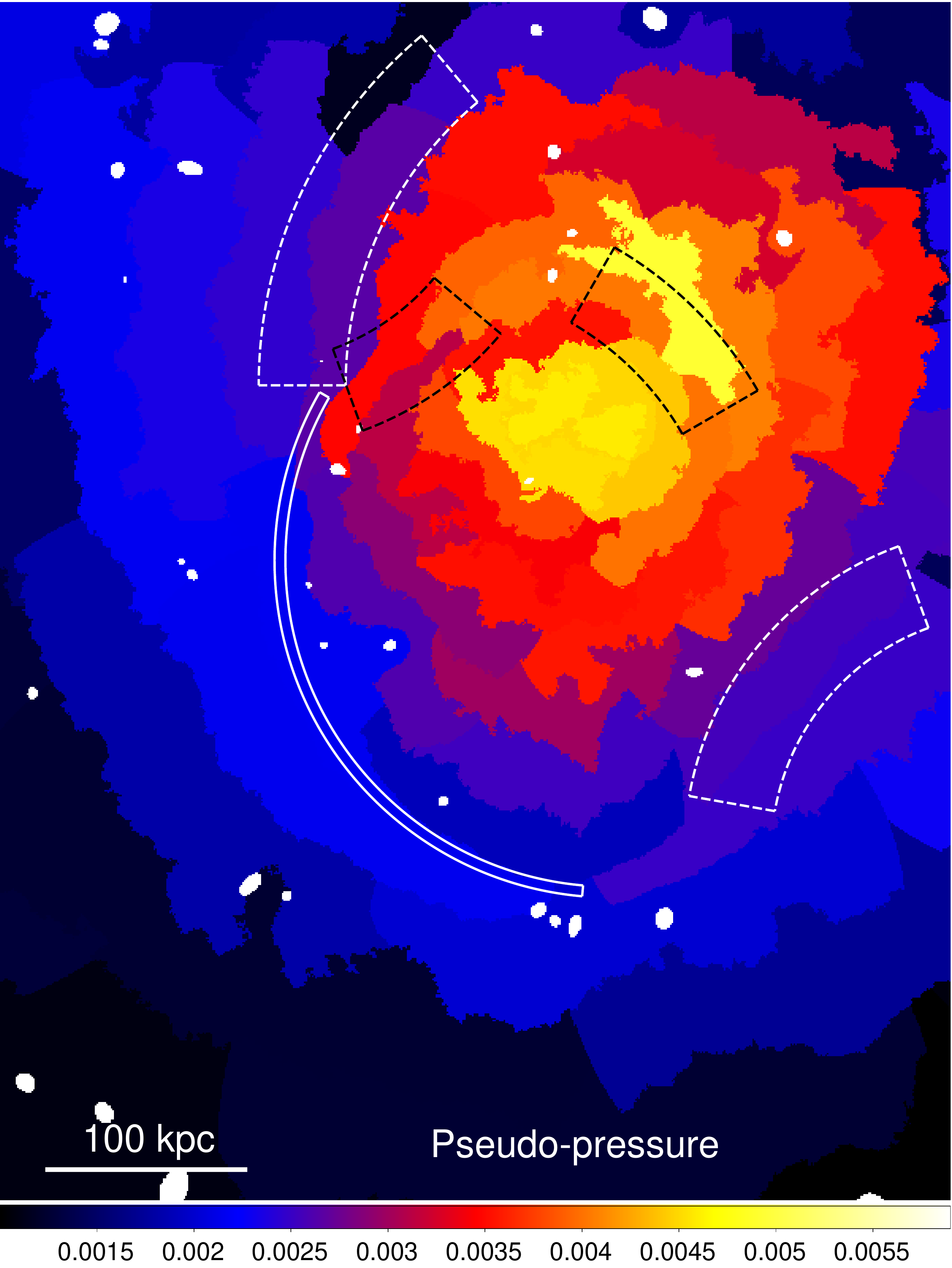}
 \end{minipage}
 \begin{minipage}{0.33\hsize}
  \centering
  \includegraphics[width=2.2in]{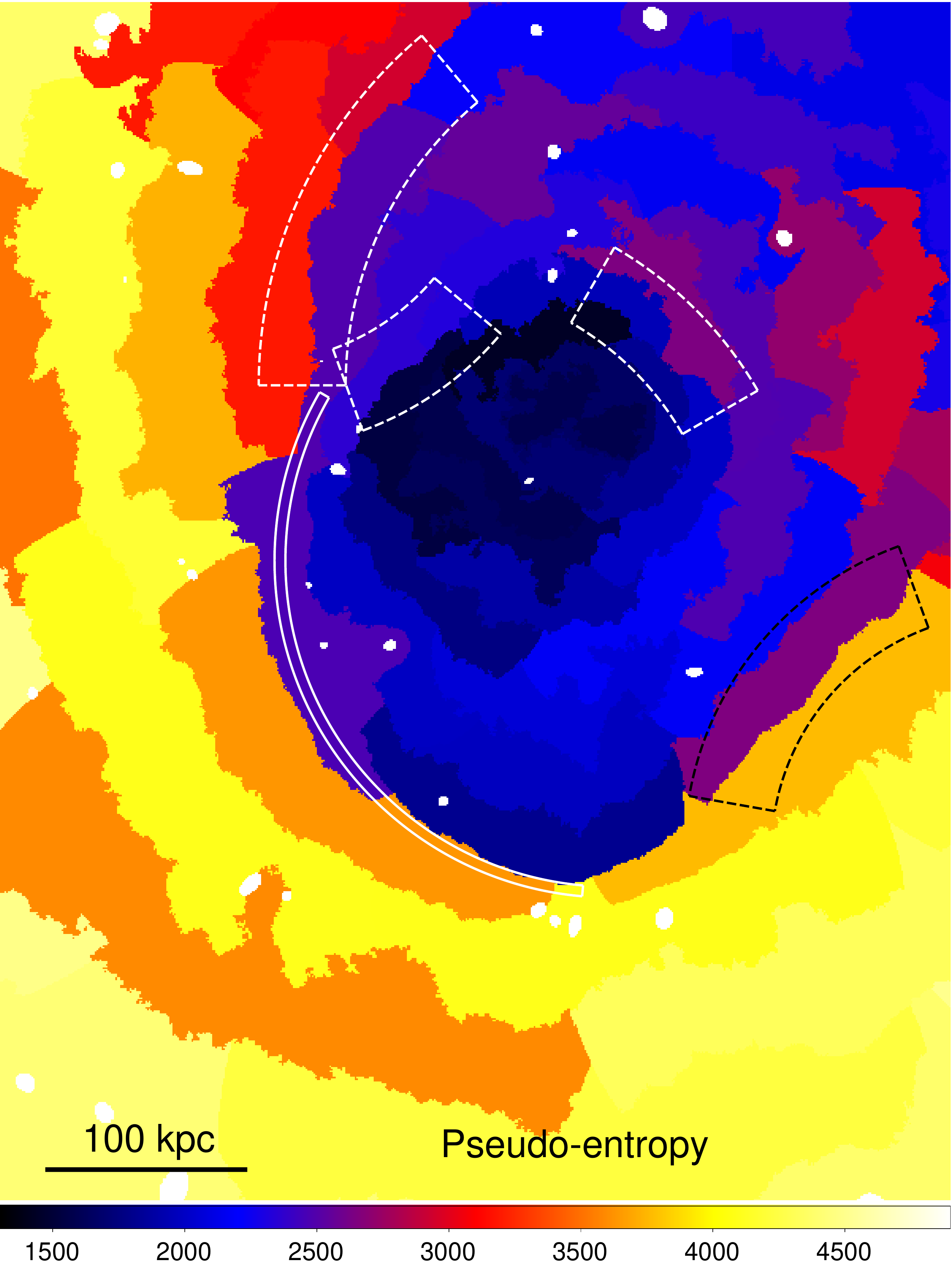}
 \end{minipage}
 \cprotect\caption[]{Projected thermodynamic maps. {\it Left:} projected temperature map in units of keV. {\it Middle:} pseudo-pressure map. {\it Right:} pseudo-entropy map. The position of the cold front is marked by the thin solid white partial annuli. The positions of other features are indicated by the dotted partial annuli (see also Fig.~\ref{img:flatimage}). The white circles/ellipses are the positions of point sources which are identified using the \verb+wavdetect+ tool.}
 \label{img:thermomaps}
\end{figure*}

\begin{figure}
  \centering
  \includegraphics[width=3.0in]{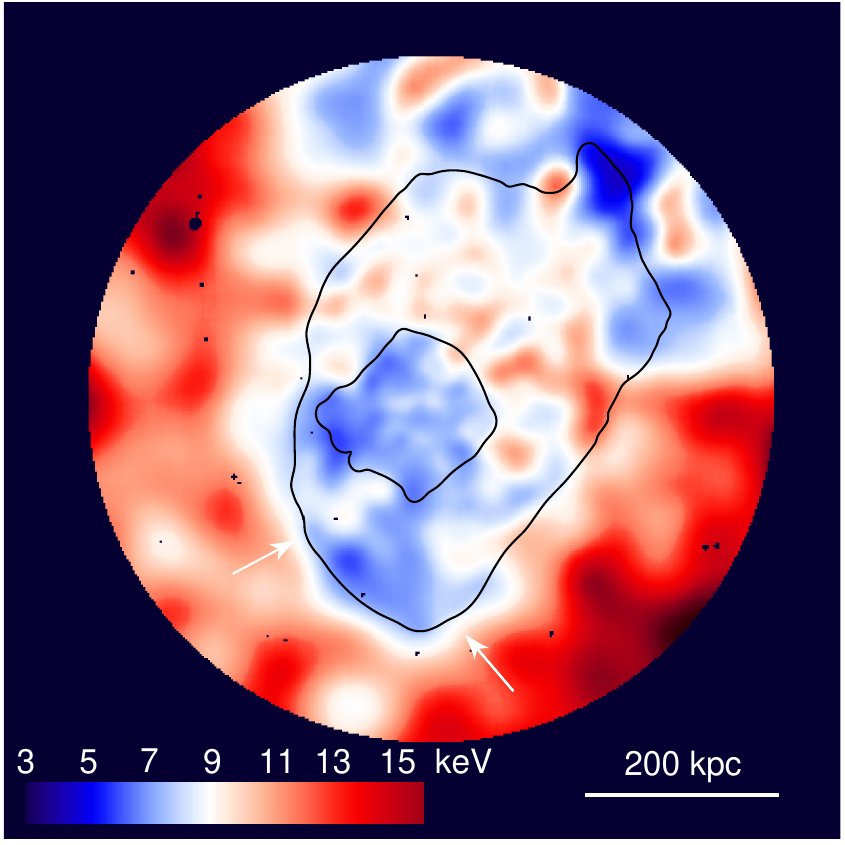}
 \caption[]{{\it Chandra} projected gas temperature map derived using a different method (reproduced from Wang 2019). Black contours show X-ray surface brightness (the outer contour is at the cold front). Arrow on left points to a hot spot inside the front (a circle in Fig.~\ref{img:flatimage}; discussed in Sec.~\ref{sec:pocket}). Arrow on right points to the region of apparent KH instabilities including the `finger', discussed in Sec.~\ref{sec:khi}.}
 \label{fig:qian_tmap}
\end{figure}

All the spectral fits are performed using {\small XSPEC} \citep[version 12.10.0c;][]{arnaud96}. We modelled the spectra using the \verb+TBabs*apec+ model with the redshift and Fe abundance fixed to 0.0557 and 0.3\,solar, respectively. The hydrogen column density was set to 11.2$\times$10$^{20}$\,cm$^{-2}$ determined by \citet{willingale13}; this is because Abell~2319 is located in a region of relatively high absorption and \citet{ghirardini18} pointed out that the $Chandra$ spectra are better modelled using the column density value corrected for molecular hydrogen than using that determined by the LAB (Leiden/Argentine/Bonn) radio HI survey \citep{kalberla05}.

We also created the thermodynamic maps. We used the contour binning algorithm \citep{sanders06} to divide the field of view into subregions used for spectral fitting. The signal-to-noise ratio of each bin is set to 100, corresponding to $\sim$10000 counts/bin. Using the best-fit temperature kT and normalization $\epsilon$, we calculated the pseudo-pressure $\tilde{p}=kT\sqrt{\epsilon/A}$ and pseudo-entropy $\tilde{s}=kT(\epsilon/A)^{-1/3}$ where $A$ is the area of the corresponding region measured in the unit of pixels. The resulting thermodynamic maps are shown in Fig.~\ref{img:thermomaps}.

\subsection{Surface brightness profiles of the main cold front}\label{sec:sbprof}
\begin{figure}
  \centering
  \includegraphics[width=3.4in]{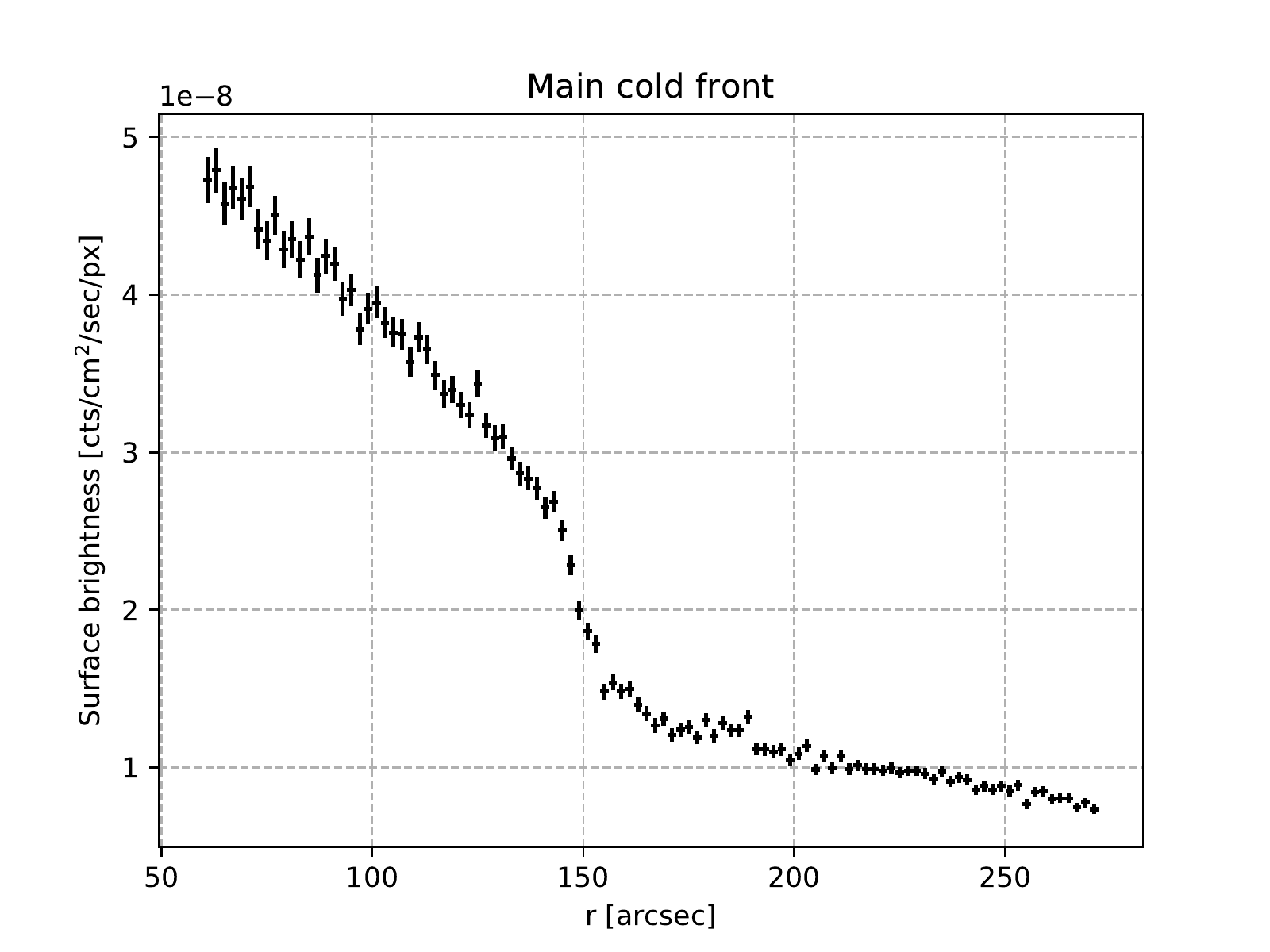}
 \caption[]{Surface brightness profile of the main front. The profile is extracted from the azimuthal range where the front is most prominent (see the dashed partial annulus to southeast in Fig.~\ref{img:flatimage}).}
 \label{img:prof_main}
\end{figure}

Fig.~\ref{img:prof_main} shows the surface brightness profile of the main cold front, extracted in the entire azimuthal range where the front is most prominent ($150^\circ-270^\circ$, see Fig.~\ref{img:flatimage}). In order to investigate the detailed azimuthal variation of the front properties, we extracted surface brightness profiles across the cold front with opening angles of 15$^\circ$ and 5$^\circ$. The directions from which the 15$^\circ$ surface brightness profiles are extracted are indicated by the white solid partial annuli shown in Fig.~\ref{img:flatimage}. The center is determined at (RA, Dec)$=$(19:21:10.2,+43:55:46.1) so that the radial directions are perpendicular to the interface. 

We modelled the observed profiles by assuming that the surface brightness is proportional to the square of a spherically symmetric density profile integrated along the line-of-sight (LOS). We used the underlying radial density profile of a broken power law with a normalization jump of $j_{12}$ at the cold front radius $r_{12}$;
\begin{equation}
 n(r) = \left\{
         \begin{array}{l}
          j_{12}n_0\left(\dfrac{r}{r_{12}}\right)^{-\alpha_1}\ \ (r \leq r_{12})\\
          n_0\left(\dfrac{r}{r_{12}}\right)^{-\alpha_2}\ \ (r_{12} < r)
         \end{array}
        \right.,\label{eq:bknpow}
\end{equation}
where $n_0$ is the overall normalization, and $\alpha_1$ and $\alpha_2$ are the power-law slopes of the density profile inside and outside the interface.

\begin{figure}
  \centering
  \includegraphics[width=3.2in]{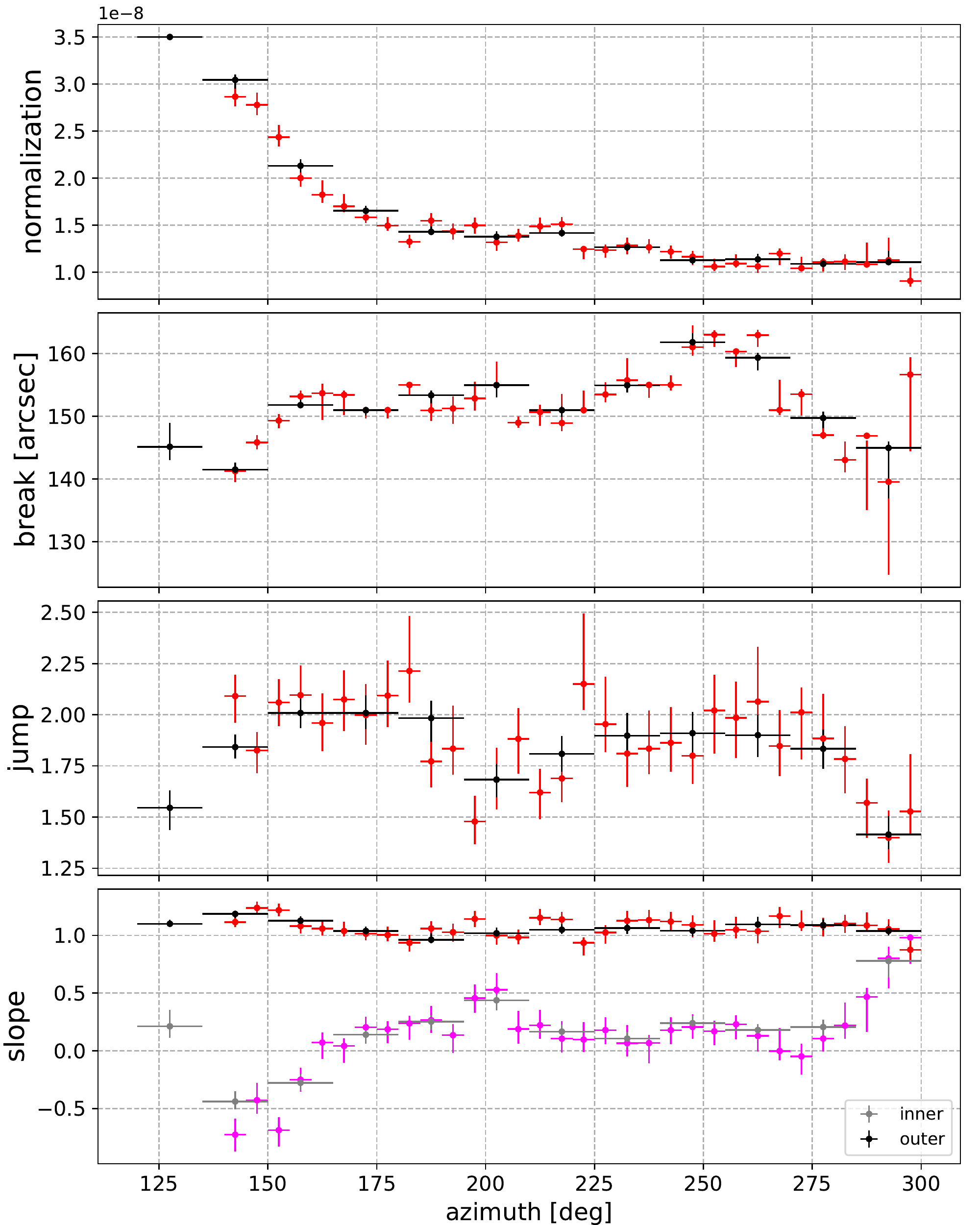}
 \caption[]{The azimuthal variations of the parameters of the main cold front within the sector shown in Fig.~\ref{img:flatimage}. The parametes of the density model, given by Eq.~\ref{eq:bknpow}, are shown in panels from top to bottom: the normalization $n_0$, the density jump radius $r_{12}$, the jump amplitude $j_{12}$, the power-law slope inside (inner)/outside (outer) the break ($\alpha_1$ and $\alpha_2$). The black/grey and the red/magenta points represent the best-fit parameters for the 15$^\circ$ (sectors in Fig.~\ref{img:flatimage}) and 5$^\circ$ profiles, respectively, counterclockwise starting from Northeast. The azimuthal range where the front is most prominent (indicated by the partial dashed annulus to southeast in Fig.~\ref{img:flatimage}) corresponds to $150^\circ-270^\circ$.}
 \label{img:azimuthal_variation}
\end{figure}

In the fitting, we defined the likelihood function to be half of the chi-square function, corresponding to the `exponential-term' of log-Gaussian, and performed maximum-likelihood fitting of the model to the observed data using the {\small emcee} Markov chain Monte Carlo (MCMC) package \citep{emcee}. The error bars are quoted based on the 16th and 84th percentiles of the samples in the marginalized distributions. Fig.~\ref{img:azimuthal_variation} shows the azimuthal variations of the best-fit parameters. The black/grey and the red/magenta points represent the best-fit parameters for the 15$^\circ$ and 5$^\circ$ profiles, respectively.

\section{Results}
\subsection{Morphological features}\label{sec:morph}
As we can see both in Fig.~\ref{img:flatimage} and Fig.~\ref{img:prof_main}, there is a clear brightness edge in the southeastern direction. The thermodynamic maps shown in Fig.~\ref{img:thermomaps} indicate that the temperature and entropy increase abruptly across the interface from the brighter to the fainter sides, while the pressure shows a continuous change. These behaviours are typical of cold fronts and consistent with the cold front observations in the literature \citep[e.g.][]{markevitch07}.

The interface is relatively smooth but shows non-monotonicity in radius. For example, the radius is locally smaller compared to neighboring azimuths around the direction indicated by the left white arrow in Fig.~\ref{img:flatimage}, and increases locally around the middle white arrow. We also see a bright `finger' of cool bright gas extending towards the west from the southernmost edge of the interface denoted by the right white arrow. It is located right at the azimuth where the cold front outline bends (the intersection of the main cold front and the edge {\it 4}). Inside the interface, we see a local depression in surface brightness as denoted by the white dashed circle. Some of these features have been pointed out also by \citet{wangphd}.

In addition to the main cold front, we found several other brightness features that have not been explicitly discussed in the literature; (i) we found that the main cold front diverges at its northern end. One of the branches extends further in the northern direction (marked as {\it 1} in Fig.~\ref{img:flatimage}), while the other bends towards the core direction (marked as {\it 3} in Fig.~\ref{img:flatimage}). (ii) we found another brightness edge to northwest from the brightness peak (marked as {\it 2} in Fig.~\ref{img:flatimage}). The main coldfront, the edge {\it 3} and the edge {\it 2} seem to be aligned on the same spiral in this order. The edge {\it 3} seems to be rather concave than convex.  (iii) we found another concave brightness edge to southwest from the core (marked as {\it 4} in Fig.~\ref{img:flatimage}). As shown in Fig.~\ref{img:thermomaps}, either the temperature ({\it 2}, {\it 3} and {\it 4}) or the entropy ({\it 1}, {\it 2} and {\it 4}) seem to exhibit abrupt changes at the interface, suggesting that all of the brightness features are cold fronts.

\subsection{Azimuthal variation of the cold front}\label{sec:cf}
As shown in Fig.~\ref{img:azimuthal_variation}, all the density profile parameters of the main cold front show significant azimuthal variations. The azimuthal profile of the break radii is not a monotonic function, and we can see several local maxima both in the 15$^\circ$ and 5$^\circ$ profiles around $160^\circ-170^\circ$, $200^\circ$ and $240^\circ-260^\circ$. In particular, the the break radius abruptly decreases at $\sim205^\circ$ and increases at $\sim245^\circ$, which are consistent with the morphological observation in the previous section.

The azimuthal profile of the density jump is mostly flat between $140^\circ$ and $280^\circ$, but shows a weakening of the jump around $200^\circ$. The overall trend is consistent with what is shown in \citet{walker16}, where the jump profile was computed using the Gaussian gradient magnitude filtering method. The azimuthal profiles of the density slope outside and inside the edge both show apparent fluctuations. The inner slope takes a local maximum around $200^\circ$, where the jump weakens. Towards the position angle of $285^\circ-300^\circ$, both the density jump and the difference in power-law slopes inside versus outside the break decrease.

\subsection{Modelling of the brightness `finger'}\label{sec:khifit}
\begin{figure}
  \centering
  \includegraphics[width=3.4in]{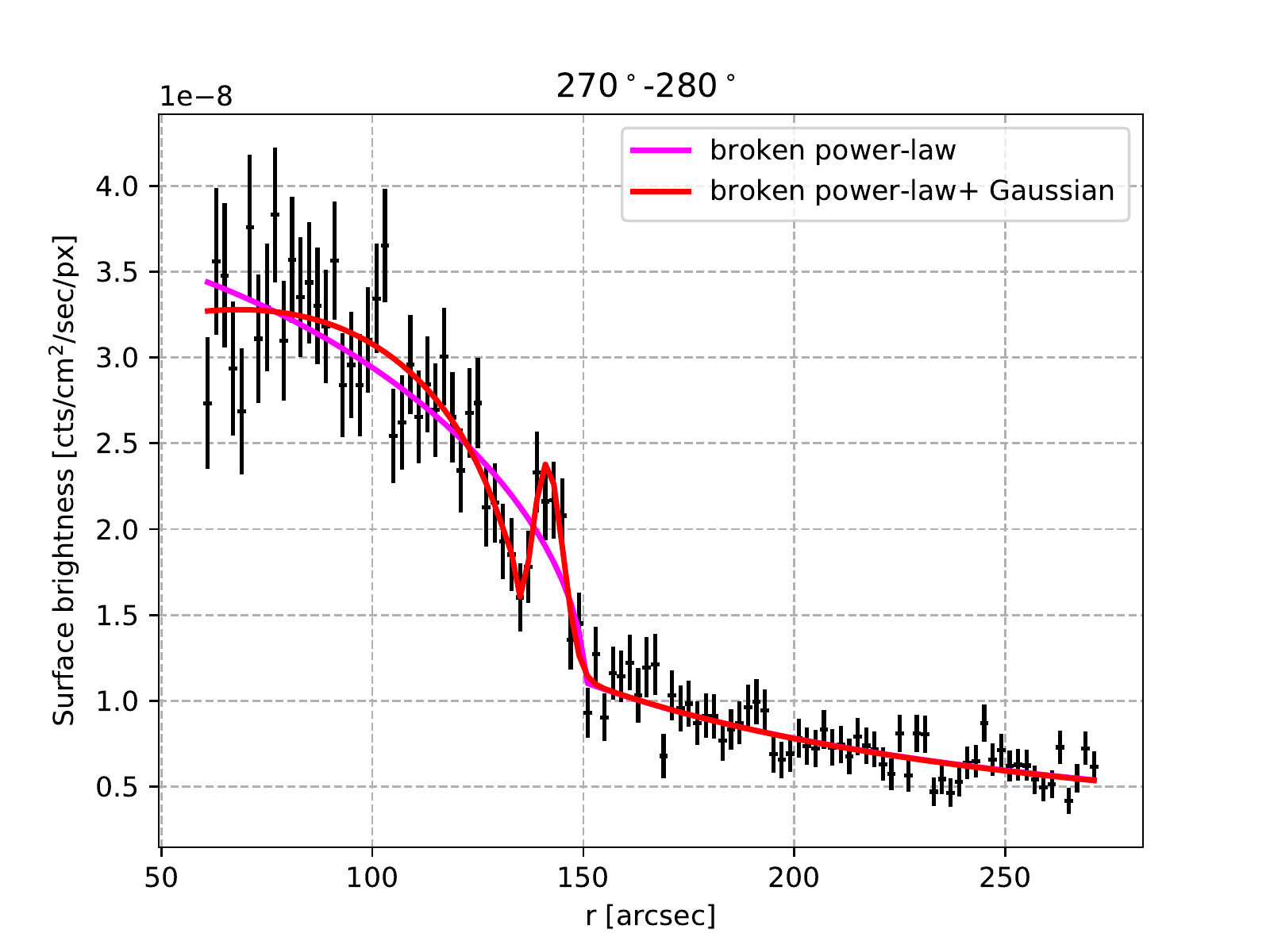}
 \caption[]{Surface brightness profile extracted in the direction across the `finger' feature. The magenta curve is the best-fitting model using the baseline model described in Section~\ref{sec:sbprof}. The red curve is the best-fitting model in which the `finger' feature is accounted for as a Gaussian component.}
 \label{img:prof_270-280}
\end{figure}

Fig.~\ref{img:prof_270-280} shows the surface brightness profile extracted in the direction across the `finger' feature. There is a coherent surface brightness excess with respect to the baseline model (magenta curve) around $r=140$\,arcsec, corresponding to the location of the `finger' in the image (Fig.~\ref{img:flatimage}).

In order to check if this is real, we added a Gaussian component that phenomenologically represents the `finger' to our baseline model described in Section~\ref{sec:sbprof}, and fit the surface brightness profile using this model. The red curve in Fig.~\ref{img:prof_270-280} shows the best-fitting surface brightness model with the `finger' feature. The fit around the feature is apparently improved, with the improvement of the fit at the significance level of 4.1$\sigma$ according to the likelihood-ratio test ($-2\Delta\ln L = 23.3$ for 3 degrees of freedom (dof))\footnote{Assuming that the number of azimuths of 12 corresponds to the number of independent random trial, the statistical significance decreases to $\sim$3.3$\sigma$ by the look-elsewhere effect.}. The reduced chi-square values with respect to the best-fit models are $\chi^2/$dof=122.1/101 and 98.8/98 without and with the Gaussian component, respectively.

\subsection{Modelling of other features}\label{sec:sbothers}

\begin{figure}
  \centering
  \includegraphics[width=3.0in]{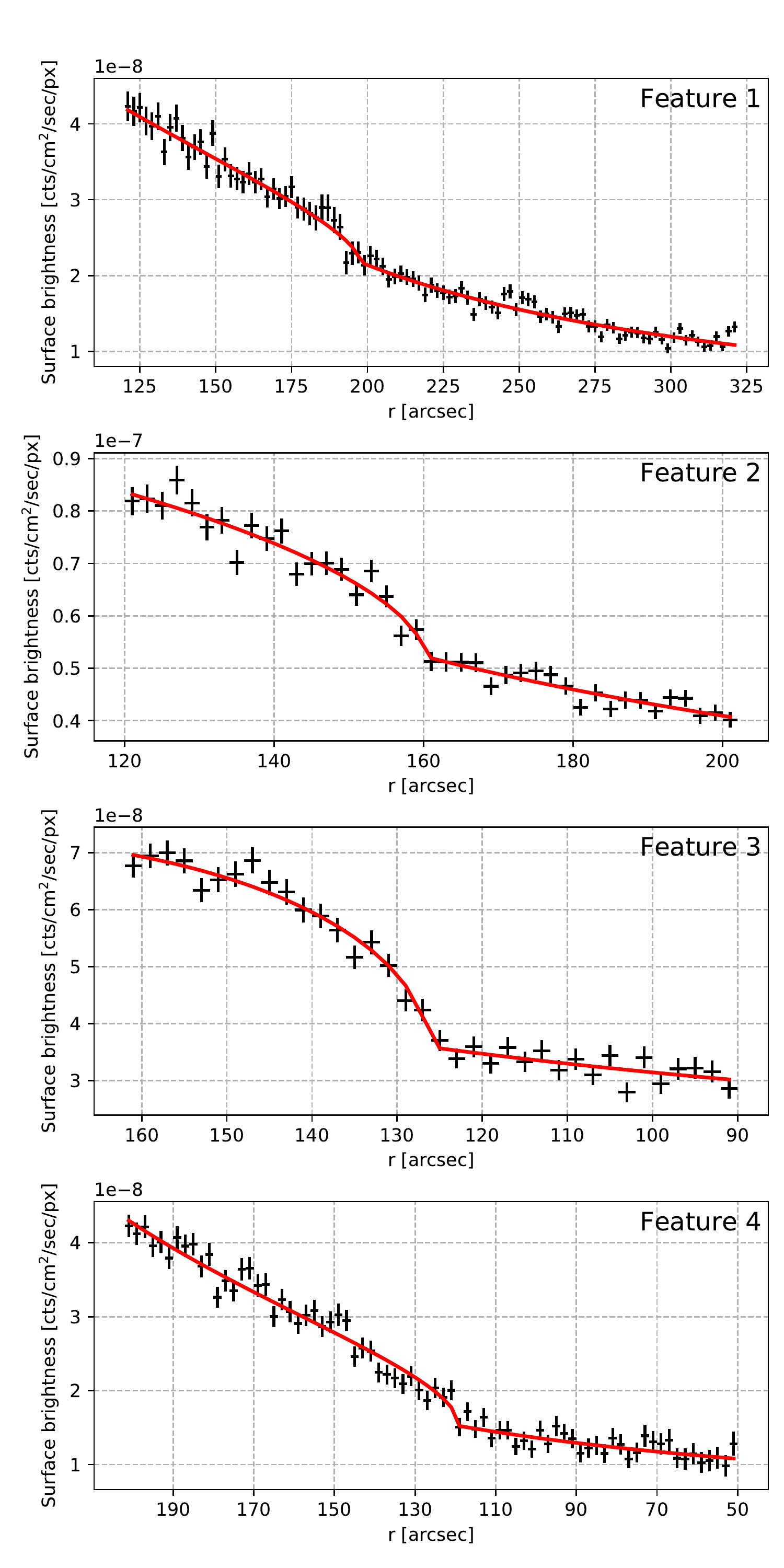}
 \caption[]{Surface brightness profiles extracted in the direction marked as {\it 1}, {\it 2}, {\it 3}, and {\it 4} in Fig.~\ref{img:flatimage} (panels from top to bottom). Note features {\it 3} and {\it 4} are concave, so the radial axis direction is reversed. The red curves are the best-fit models described in Section~\ref{sec:sbprof}. }
 \label{img:prof_others}
\end{figure}

 Fig.~\ref{img:prof_others} shows the profiles extracted in the directions marked as {\it 1}, {\it 2}, {\it 3}, and {\it 4} in Fig.~\ref{img:flatimage}, on which the respective best-fit baseline models are plotted. The centers are chosen to be at (RA, Dec)$=$(19:21:05.8,+43:57:06.7), (19:21:18.4,+43:55:34.3), (19:21:25.6,+43:59:06.8) and (19:20:53.2,+43:53:31.6), for the directions {\it 1}, {\it 2}, {\it 3}, and {\it 4}, respectively. We also modelled the profile using a model without break (simple power-law), and found that all the profiles prefer the model including the break with high significance ($-2\Delta\ln L$=81.1, 61.4, 146.0, and 264.0 for 3 degrees of freedom, for the directions {\it 1}, {\it 2}, {\it 3}, and {\it 4}, respectively). The broken power-law density model describes all the profiles well ($\chi^2/$dof=108.9/96, 43.9/36, 32.0/31, and 75.4/71 for the directions {\it 1}, {\it 2}, {\it 3}, and {\it 4}, respectively).

\section{Discussion}

\subsection{Origin of the cold fronts and the lack of the cool core}\label{sec:origin}

Cold fronts are contact discontinuities in the cluster gas that arise when bulk gas motions bring parcels of gas with different specific entropy in direct contact. The gas pressure is approximately continuous across such a front (with possible differences caused by gas velocities), while both the density and temperature experience sharp jumps. These features and their origin are discussed in \cite{markevitch07}. Cold fronts can roughly be classified into two categories\footnote{We do not consider here a separate phenomenon in which a cold front can arise from interaction of shocks \citep{birnboim10}.}. In the first one, the contact discontinuity separates the gases from different subclusters, when they collide and a dense core of one or both is stripped by ram pressure of the other cluster's gas (`stripping' cold fronts). The clearest examples of such fronts are seen in the Bullet cluster \citep{markevitch02} and the galaxy NGC~1404 falling into the Fornax cluster \citep{su17a}. The other type of cold fronts separates the lower-entropy gas from the same cluster that has been displaced from its equilibrium location at the bottom of the gravitational well, from the higher-entropy gas it encounters at the larger cluster radii. Such displacement can be caused by a merger (\cite{ascasibar06,roediger13a}; a push from a passing shock front can be sufficient, \cite{churazov03}) and sets off long-lasting back-and-forth `sloshing' of the dense core gas, producing the characteristic series of concentric cold fronts. If the disturbance imparted a nonzero angular momentum on the gas core, multiple concentric fronts could arrange into an apparent spiral, such as those seen in many clusters with cool cores \citep[e.g.,][]{churazov03,clarke04a,blanton11,osullivan12,rossetti13,ghizzardi14,ichinohe15,sanders16,ueda19,ueda20}.

\begin{figure}
  \centering
  \includegraphics[width=3.0in]{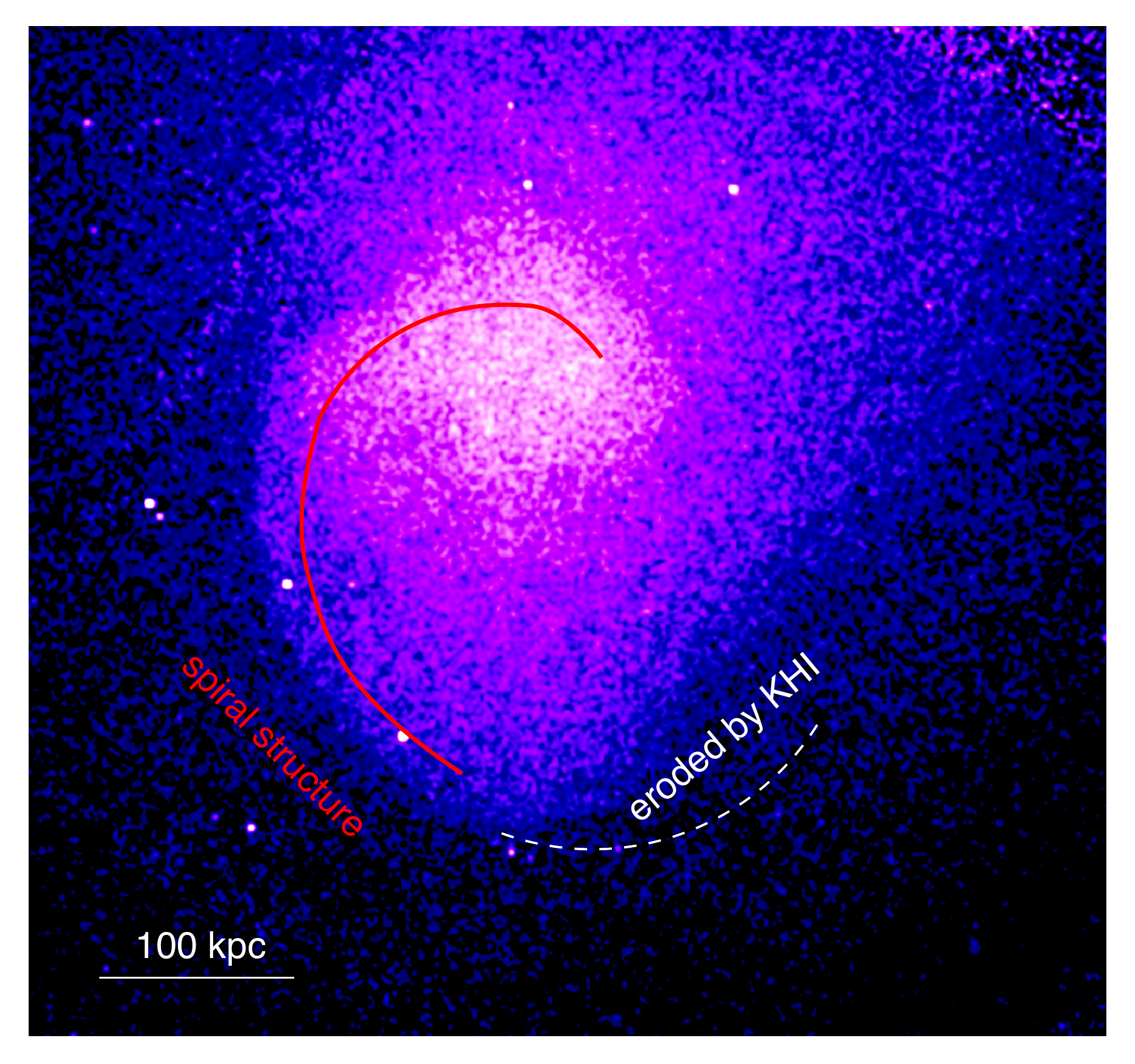}
 \caption[]{The fronts labeled {\it 2}, {\it 3} and `main' in Fig.~\ref{img:flatimage} form an approximate spiral pattern (schematically shown here by red solid line; the line is offset inward from the fronts for clarity), which is characteristic of core gas sloshing. The possible origin of the southwest concave feature is KHI developing at this position of the cold front and destroying the sharp contact discontinuity, which otherwise would continue as schematically shown by dashed line.}
 \label{fig:eroded}
\end{figure}

Note that sloshing fronts do not need a cool core --- it is the gradient of the specific entropy, rather than low temperature, that is needed to produce a contact discontinuity when the central gas is lifted up (see, e.g., Fig.~12 in \cite{ascasibar06}, which shows the core of an isothermal cluster forming a front). Indeed, Abell~2142, a `warm core' cluster based on its central entropy (\cite{cavagnolo09}), exhibits a spiral of at least four sloshing fronts (\cite{markevitch00,rossetti13,wang18}). With a central specific entropy of 270\,keV\,cm$^{2}$, Abell~2319 does not have a cool core at present, but its entropy increases with radius as in most other clusters \citep{cavagnolo09}. The fronts shown in Fig.~\ref{img:flatimage} (labeled {\it 2}, {\it 3}, `main') form a spiral pattern as schematically shown in Fig.~\ref{fig:eroded}, which indicates sloshing of the core gas. On the larger scale, the cluster has a cool X-ray extension to the northwest (discussed in \cite{markevitch96,ohara04,govoni04,ghirardini18}), which indicates a merger with at least some velocity component in the plane of the sky. This merger can be the cause of the core disturbance that set off the observed sloshing.

If the plane of sloshing motions is oriented along the line of sight, we should see disconnected, staggered surface brightness edges on the opposite sides of the cluster, while if the motions occur close to the plane of the sky, the edges should form a spiral (\cite{roediger11,zuhone18}; Fig.~19 in \cite{ascasibar06}). As shown in Figs.~\ref{img:flatimage} and \ref{fig:eroded}, the cluster is surrounded on nearly all sides by surface brightness edges, leading us to believe the latter orientation is more likely for the case of Abell~2319. This interpretation is also supported by radio observations \citep{storm15}; as we will discuss in detail in Section~\ref{sec:agn}, the radio emission consists of two components, and one of them could be interpreted as a rather bright minihalo. Although the exact origin of minihalos is still debated, they have a clear association with cool core clusters and seem to often be generated or at least reaccelerated by sloshing \citep{zuhone13b,gitti15,giacintucci17,richard-laferriere20}.

On the other hand, optical spectroscopy \citep{oegerle95} reveals a subcluster with a high LOS velocity ($>$1000\,km\,s$^{-1}$), located $\sim$10\,arcmin ($\sim$650\,kpc) in projection northwest of the core, at the end of the X-ray extension. Whether or not the gas moves with that velocity is unclear, but the 3D merger geometry probably includes an out of the sky plane velocity component, or perhaps more than one infalling subcluster. The spiral front morphology and its sharpness\footnote{For a stripping cold front seen from a small inclination angle, we would not expect the existence of such a sharp edge. This is because the edges of a stripping cold front are more likely to be smeared with the ambient medium due to hydrodynamic instabilities than the tip of the cold front \citep[see Fig.~7 of][]{roediger15b}.} argues against a stripping cold front. The spiral also strongly suggests sloshing in the sky plane, though we cannot rule out sloshing perpendicular to the sky plane (and some additional disturbance) without measuring the LOS velocities of the gas inside the fronts, which will become possible with the launch of {\it XRISM} \citep{tashiro20}. Although dedicated numerical simulations are required to prove the origin of the cold front and of the detected features, in the following sections, we will assume that the core gas is sloshing near the sky plane, while we note, in any case, that most of the discussion presented below remains valid both for a stripping as well as for a sloshing origin of the front.

The absence of a cool core in Abell~2319 is interesting. \citet{rossetti10} found that the brightest cluster galaxy (BCG) is located in a region of relatively low entropy compared to the surrounding ICM and proposed that Abell~2319 could in the past have contained a cool core. It could have been disrupted by the very sloshing that we see --- indeed, simulations by \citep{zuhone10} show that sloshing can facilitate the heat inflow into the core via mixing with the hot outer gas. A possibly more extreme example of this phenomenon is Abell~1763, where \citet{douglass18} found a large sloshing spiral (along with a merging subcluster that caused it) and a high central entropy (215\,keV\,cm$^{2}$, \citealt[][]{cavagnolo09}) characteristic of non-cool-core clusters.

\subsection{Kelvin-Helmholtz instability}
\label{sec:khi}

As we discussed in Section~\ref{sec:khifit}, the radii of the interface fluctuate with position angle. As the azimuthal profile of the break radius has multiple minima/maxima, we can deduce that the fluctuation scale is smaller than the opening angle of the entire cold front. The emergence of sub-opening-angle scale variation of the break radii is routinely seen in numerical simulations of sloshing cold fronts where KHIs develop because of gas shear \citep[e.g.,][]{roediger13a}. Observationally, the existence of the sub-opening-angle scale variation due to developing KHIs has also been indicated in the cold front in Abell~3667 by \citet{ichinohe17}.

In addition to the cold front fluctuations, we also found a `finger' feature at the position angle of $270^\circ-280^\circ$ with moderate significance. The location of the `finger' is right at the intersection of the edge {\it 4} and the main cold front, where the cold front outline bends. This feature is morphologically similar to well-developed Kelvin-Helmholtz instabilities seen in numerical simulations \citep{zuhone11,roediger13a}. It has been shown by numerical simulations and observations that a multiple-edge structure of the surface brightness profile appears when developing KHIs exist \citep{roediger13a,su17a,ichinohe17,ichinohe19a} because of the KHI eddies seen in projection. As shown in Fig.~\ref{img:prof_270-280}, we see the coherent increase in surface brightness instead of multiple edges. We suggest that this is also a projected KHI eddy, detached from the main cold front due to the bend of the cold front outline. Therefore, we think that this `finger' feature is naturally explainable with the interpretation that it is a developed KHI eddy. Note that the stripped tail of a member galaxy might be another interpretation of the `finger', but no galaxies are found in the NASA/IPAC Extragalactic Database (NED)\footnote{The NASA/IPAC Extragalactic Database (NED) is funded by the National Aeronautics and Space Administration and operated by the California Institute of Technology.} at the corresponding location.

\subsubsection{Upper limit of the ICM effective viscosity}\label{sec:viscosity}
Based on the existence (or absence) of KHI, we can derive implications on ICM microphysics. When there is a shear flow between two fluid components, KHIs start to develop. If the gas is inviscid and incompressible, all the scales show exponential development. However, finite viscosity suppresses the development of instabilities whose length scale $\lambda$ is below a certain critical value. This condition can be expressed using the critical Reynolds number $\mr{Re}_\mr{crit}$;
\begin{equation}
 \mr{Re} = \dfrac{\rho\lambda V}{\mu} < \mr{Re}_\mr{crit} \sim a \sqrt{\Delta},\label{eq:visc}
\end{equation}
where $V$ is the shear strength, and $\rho$ and $\mu$ are the density and viscosity on the high-viscosity side, respectively. $\Delta$ is calculated using the densities of gas on the two sides of the interface $\rho_1$ and $\rho_2$; $\Delta = (\rho_1+\rho_2)^2/\rho_1\rho_2$, and $a$ is a numerical coefficient around dozens \citep{roediger13b,chandrasekhar61}.

Conversely, if the KHI of the length scale of $\lambda$ exists, it indicates that the ICM viscosity should satisfy the condition
\begin{equation}
\mu<\dfrac{\rho\lambda V}{\mr{Re}_\mr{crit}}\sim\dfrac{\rho\lambda V}{a\sqrt{\Delta}},\label{eq:viscul}.
\end{equation}
Therefore, when one observes instabilities of scale $\lambda$, one can obtain the upper limit on the ICM viscosity using the gas densities and the shear strength.

\begin{figure}
  \includegraphics[width=1.6in]{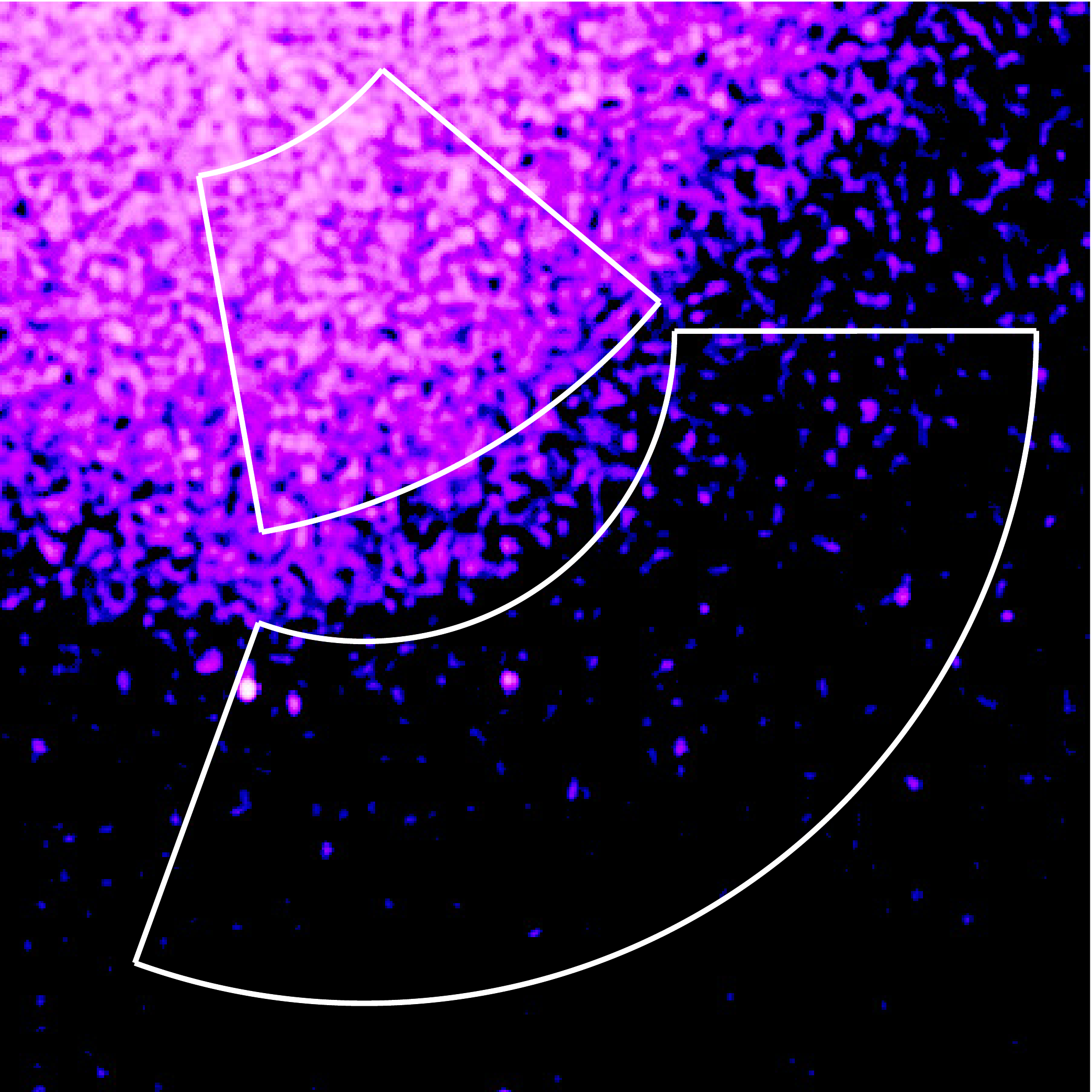}
  \includegraphics[width=1.6in]{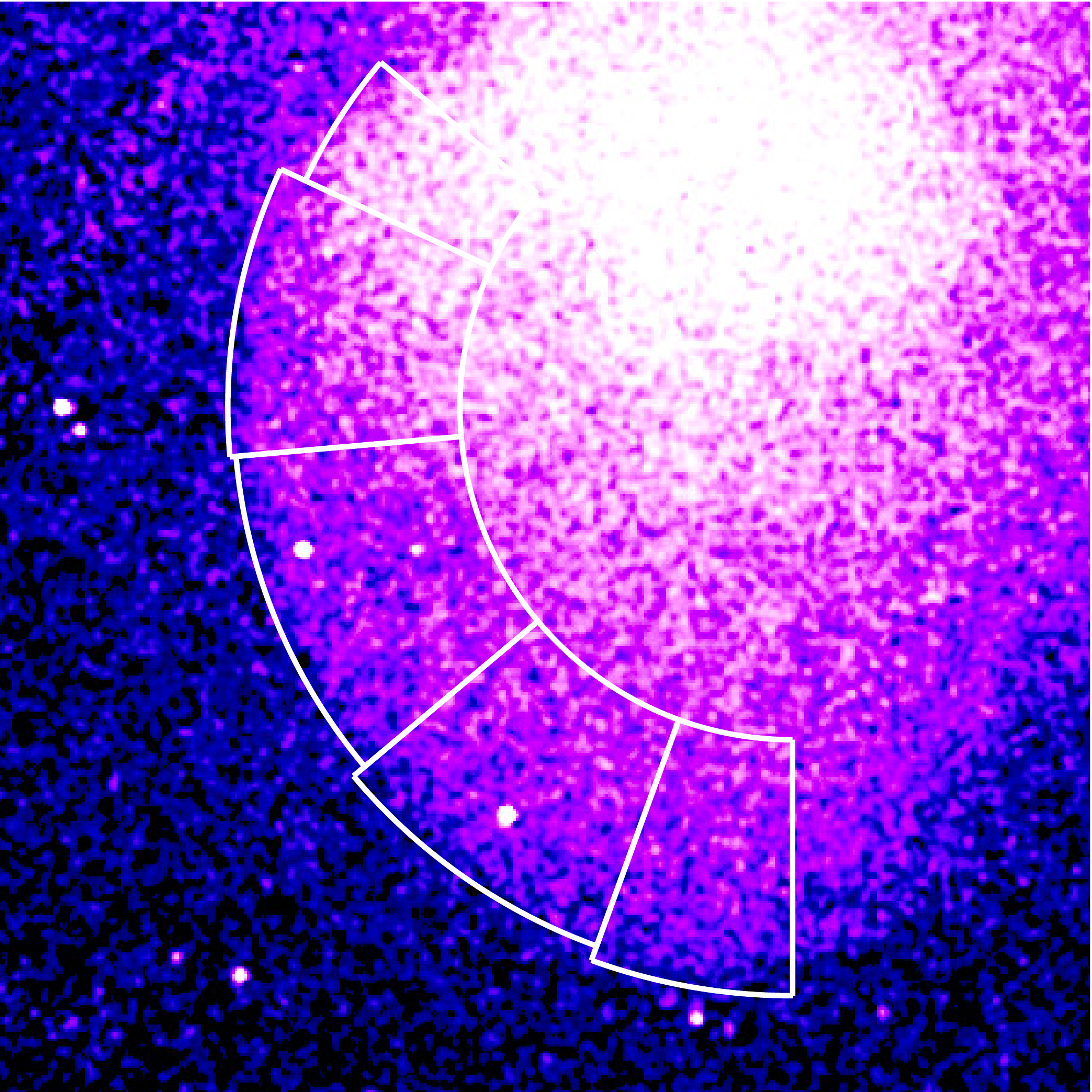}
 \caption[]{{\it Left:} Same as Fig.~\ref{img:flatimage}, close-up on the KHI candidate. The color bar is changed to visually emphasize the `finger' feature. The overlaid regions are used for spectral analysis in Section~\ref{sec:khi}. {\it Right:} Same as Fig.~\ref{img:flatimage}. The overlaid regions are used for spectral analysis in Section~\ref{sec:pocket}.}
 \label{img:flatimage_khi} \label{img:flatimage_pocket}
\end{figure}

\begin{table*}
 \centering
 \caption{Best-fitting thermodynamic parameters for the region shown in Fig.~\ref{img:flatimage_khi}.}
 \label{tbl:specfit_khi}
 \begin{tabular}{cccc}
  \hline
  Region & Temperature (keV) & Fe abundance (solar) & Electron density ($10^{-3}\mr{cm}^{-3}$)\\
  \hline
  Inner & 9.1$\pm$0.5 & 0.59$\pm$0.12 & $8.8\pm 0.1\times (L_\mr{in}/100\,\mr{kpc})^{-1/2}$ \\
  Outer & 12.7$\pm$1.0 & 0.53$^{+0.19}_{-0.18}$ & $4.9\pm0.1\times (L_\mr{out}/100\,\mr{kpc})^{-1/2}$ \\
  \hline
 \end{tabular}
\end{table*}

In order to estimate the gas properties, we extracted spectra from the regions shown in Fig.~\ref{img:flatimage_khi} left, where the upper left and lower right partial annuli correspond to the gas inside (Inner) and outside (Outer) the interface, respectively. Note that we deliberately avoid the KHI region because the KHI would mix the gas from both sides making the interpretation difficult. The best-fit parameters are shown in Table~\ref{tbl:specfit_khi}, where $L_\mr{in}$ and $L_\mr{out}$ are the LOS depth of the gas contributing to the spectra extracted from the Inner and Outer regions, respectively.

These gas properties are combined with Eq.~\ref{eq:viscul} to obtain the upper limit on the viscosity;
\begin{equation}
\begin{split}
\mu&\lesssim\dfrac{\rho\lambda V}{a\sqrt{\Delta}}\\&\sim5900\,\mr{g}\,\mr{cm}^{-1}\,\mr{s}^{-1}\left(\dfrac{n_\mr{out}}{3\times10^{-3}\,\mr{cm^{-3}}}\right)\left(\dfrac{\lambda}{100\,\mr{kpc}}\right)\left(\dfrac{V}{1600\,\mr{km}\,\mr{s}^{-1}}\right)\\
&\times\left(\dfrac{a}{10}\right)^{-1}\left(\dfrac{1/\sqrt{\Delta}}{0.4}\right)
\label{eq:viscul_num},
\end{split}
\end{equation}
where $n_\mr{out}$ is the electron density outside the interface derived in \citet{ghirardini18}. The choices of the rather conservative fiducial values in deriving Eq.~\ref{eq:viscul_num} are explained below.

(i) $\lambda$ is the scale of the KHI. Based on the `eroded front' scenario discussed in detail in Section~\ref{sec:eroded}, assuming that the pitch of the {\it feathery} structures represents the wavelength of the scale that can grow, $\lambda\sim$100\,kpc can be inferred. The temperature map in Fig.~\ref{fig:qian_tmap} exhibits multiple eddy-like structures of cool gas at the `eroded' region and the pitch of the cool gas components is similarly about 100\,kpc. The current data quality precludes the structures of smaller scales from being detected and it is possible that the existing scales may even be smaller, but in any case, $\sim$100\,kpc seems to be a reasonable choice for the conservative estimation of $\lambda$. (ii) $V$ is the relative shear velocity of the gases on two sides of the front. Since gas sloshing is subsonic, the Mach number of the shear is expected to be well below the sound speed \citep{markevitch01,wang18}. As we do not observe shock features in the image, we simply adopted the current fiducial value of $V\sim1600\,\mr{km}\,\mr{s}^{-1}$, the sound speed for the cooler side temperature (9.1\,keV, see Table~\ref{tbl:specfit_khi}). (iii) $a$ is the numerical coefficient and several estimations ($a\sim10$, 16 and 64) are presented by \citet{roediger13b}. The most conservative choice is $a=10$ and we took this value. Note that it is pointed out in \citet{ichinohe17} that the values 10 and 16 may be too conservative because some simulations with the Reynolds number above these values show the suppression of instabilities \citep[e.g., Fig.~8 in][]{roediger15b}. (iv) The $1/\sqrt{\Delta}$ factor in Eq.~\ref{eq:viscul} depends weakly on the actual geometry and takes a value around $\sim$0.4 for $L_\mr{out}/L_\mr{in}=1-10$.

The isotropic, temperature-dependent Spitzer viscosity is
\begin{equation}
 \mu_\mr{Sp} = 15000\,\mr{g}\,\mr{cm}^{-1}\,\mr{s}^{-1}\left(\dfrac{kT}{12.7~\mr{keV}}\right)^{5/2},\label{eq:spitzer}
\end{equation}
where the Coulomb logarithm $\ln\Lambda=40$ is assumed \citep{spitzer,sarazin86,roediger13a}. We can safely reach the conclusion that the ICM viscosity is suppressed by at least factor 2.5 below the full Spitzer value.

Recently, several attempts have been made to infer the ICM viscosity by this methodology. Using the merger cold front in Abell~3667, \citet{ichinohe17} estimated the upper limit of the ICM effective viscosity to be $\mu\lesssim 200\,\mr{g}\,\mr{cm}^{-1}\,\mr{s}^{-1}$. Similarly, using the merger cold front in NGC~1404 in the Fornax cluster, \citep{su17a} put an upper limit at 5\% of the Spitzer value.

The estimation of the upper limit of the ICM viscosity using sloshing cold fronts (especially sloshing in the plane of the sky) has been difficult due to the difficulty of estimating the shear velocity perpendicular to the LOS direction. Estimates based on the comparison of the observation to the tailored simulation have been made \citep{roediger13a,werner16a}. Using the sloshing cold front in Abell~2142, which exhibits multiple KH eddies, and the gas velocity estimate based on centripetal acceleration, \citet{wang18} derived the first quantitative upper limit, $\mu<1/5\,\mu_{\rm Sp}$.

Our result is consistent with the picure presented in the literature that the ICM viscosity is suppressed from the full Spitzer viscosity. The reason that we were able to derive a meaningful upper limit using very conservative parameters, is the high temperature ($\gtrsim$10\,keV) of the system. We used the sound speed, which depends on the square-root of the temperature, for the shear velocity $V$ (Eq.~\ref{eq:viscul}), while the full Spitzer viscosity is proportional to $kT^{5/2}$ (Eq.~\ref{eq:spitzer}). The viscosity suppression fraction is proportional to the inverse of the temperature squared; $\mu/\mu_\mr{sp}\propto kT^{-2}$, and therefore higher temperature systems are more advantageous.

\subsection{Split of the main cold front}
As shown in Fig.~\ref{img:flatimage}, the northern end of the main cold front appears to split in two -- one extends rather along the way of the main cold front (marked as {\it 1}), and the other bends sharply to inside the interface (marked as {\it 3}). A similar split of the cold front is found at $\sim$730\,kpc from the Perseus cluster core \citep{walker18}. Several observational/simulation studies have suggested the existence of such a double-layered structure of cold fronts due to KHIs. These instabilities can develop regardless of the cold front origins: see e.g. \citet{roediger13a,ichinohe19a} for sloshing cold fronts, and e.g. \citet{su17a,ichinohe17} for stripping cold fronts. If one thinks of the cold front as a 3D surface, this double edge may be a projection of an inward depression in the front surface created by KHI. We will also discuss a more exotic possibility below.

\begin{figure}
  \centering
  \includegraphics[width=3.0in]{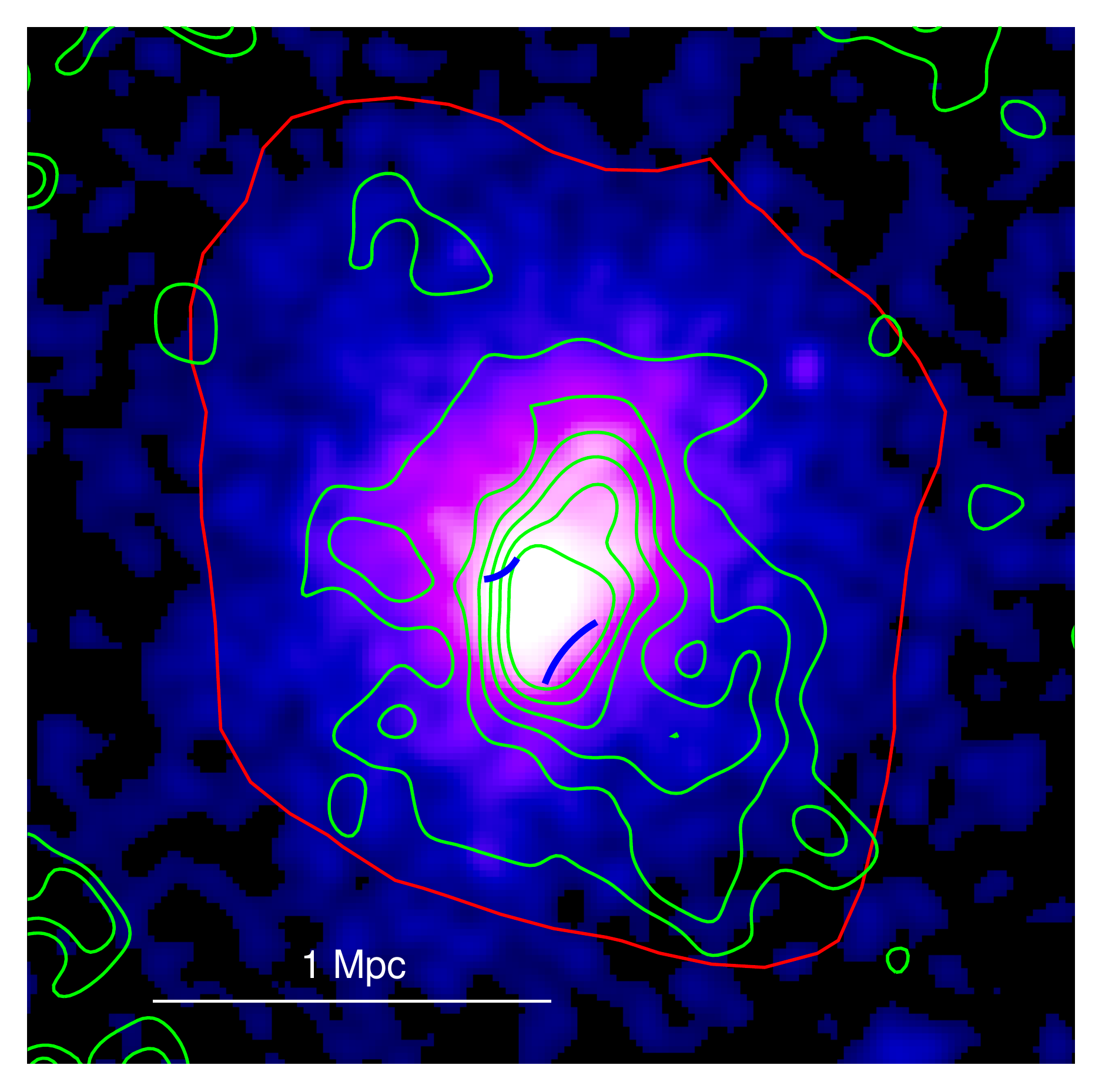}
 \caption[]{A wider-field archival {\it ROSAT} PSPC image of Abell~2319 (0.5--2\,keV), with two concave segments of the cold front seen by {\it Chandra} (the edges {\it 3} and {\it 4} in Fig.~\ref{img:flatimage}) marked by blue arcs near the center. Contours overlay diffuse radio emission at 1.4\,GHz from {\it VLA} (green) and {\it GBT} (red), reproduced from \citet{storm15}.}
 \label{fig:radio}
\end{figure}

\subsection{Southwest concave discontinuity}
Southwest of the center, the main cold front continues as an interesting, much less sharp, concave surface brightness discontinuity (the edge {\it 4} in Fig.~\ref{img:flatimage}). Below we discuss two possible scenarios for it.

\subsubsection{Erosion of the front by KHI}\label{sec:eroded}
One possible cause of this feature is the gas dynamics related to core sloshing, similar to the interpretation suggested by \citet{werner16b} for a concave discontinuity in the Ophiuchus cluster. Hydrodynamic simulations suggest that such deformations of the cold front shape can arise during core sloshing \citep[see e.g., Fig.~22 in][]{ascasibar06}. The gas flow tangential to the front surface should result in KHIs, which at some locations should grow fast enough to disrupt the discontinuity entirely\footnote{Note that the erosion of cold fronts may occur also in stripping, and thus the exact origin of the cold front does not affect for this discussion.}. The southwest concave structure may be such a region of the front eroded by KHI, as schematically depicted with the dashed line in Fig.~\ref{fig:eroded}.

Although the surface brightness profile (Fig.~\ref{img:prof_others} bottom) and the thermodynamic maps (Fig.~\ref{img:thermomaps}) both indicate that the gas properties are different inside and outside this edge, neither the profile nor the image (Fig.~\ref{img:flatimage}) show sharp features. A close inspection of the X-ray image shows wavy {\it feathery} structure of the discontinuity, similar to the Perseus cluster cold front \citep{ichinohe19a}. The `finger' discussed in Section~\ref{sec:khi} is the most prominent of those {\it feathers}. This region may have large KH rolls mixing the gases on two sides of the former discontinuity. In projection, they may not appear as the typical simulated KHI rolls \citep[e.g.,][]{roediger13a} because of their more complex 3D structure.

\subsubsection{A giant AGN outburst?}\label{sec:agn}
We also consider an interesting, but at this point rather speculative alternative possibility. A similar concave X-ray feature near the core of the Ophiuchus cluster, upon comparison with the radio data, turned out to be the result of the most powerful known AGN explosion \citep{giacintucci20}. AGN outbursts generate powerful jets that transport large amounts of energy from the nucleus and deposit it at a distance, typically within the cluster cool core. This process creates a pair of bubbles in the ICM, filled with relativistic particles mixed with hot ICM. In the X-ray images, these bubbles are seen as symmetric pairs of round cavities or depressions; typical examples are the Perseus cluster \citep{fabian06}, Hydra A \citep{nulsen05} and MS0735+74 \citep{mcnamara05}. Apparently, the outburst can be powerful enough for these bubbles to form outside the cool core, as in the Ophiuchus cluster \citep{giacintucci20}. In such a case, the X-ray cavity can be difficult to see in its entirety --- only its innermost segment, closest to the core, may be seen as a concave feature. However, the pair of diffuse radio lobes filling the cavity should always be present --- though, as in Ophiuchus, they may emit at very low radio frequencies.

In Fig.~\ref{fig:radio}, we show a wider-field archival X-ray image of Abell~2319 from {\it ROSAT} PSPC, with overlays of 1.4\,GHz radio maps from {\it VLA} and {\it GBT}, reproduced from \citet{storm15}. For {\it VLA}, compact sources have been removed to the degree possible. For {\it GBT}, only the lowest contour is shown. The {\it GBT} angular resolution is low, comparable to the size of the cluster cool core, but it does not miss the faint extended emission on large angular scales, because it is a full-aperture instrument. The radio emission consists of a relatively bright minihalo candidate that fills the bright core of the cluster and follows the distribution of the lowest temperature ICM, and a more extended emission, which \citet{storm15} interpret as a giant radio halo enveloping the minihalo (for review of the types of diffuse radio sources in clusters see \citealt{vanweeren19}). A similar observation was made by \citet{govoni04} using less sensitive {\it WSRT} radio data and earlier {\it Chandra} data. 

Interestingly, the {\it GBT} image reveals a faint radio extension to the NE, opposite to the SW extension seen by the {\it VLA} (it may have been missed by the {\it VLA} because of the limitations of an interferometer). Fig.~\ref{fig:radio} also marks the two concave X-ray features that we find on the opposite sides of the cluster core. It is conceivable that those features are the inner boundaries of two giant X-ray cavities, and the radio emission that extends beyond the minihalo is in fact a pair of giant radio lobes that fill those cavities, similar to the Ophiuchus lobe.

Abell~2319 is of course not completely analogous to Ophiuchus --- Ophiuchus has a dense cool core and its central galaxy has a weak radio source (AGN) that could have powered the outburst, while Abell~2319 has no cool core and its BCG does not show any radio source in the current data (there is a faint source but it is offset from the galaxy center). However, despite the powerful past outburst, the AGN in Ophiuchus is currently very weak, because core sloshing has displaced the gas density peak from it and apparently starved it of accretion material \citep{werner16b}. It is quite possible that the disturbance that caused sloshing in Abell~2319 has disrupted its cool core (Section~\ref{sec:origin}) and starved its AGN even more thoroughly. It is also possible that the past AGN outburst itself has disrupted its fuel supply.

The energy of an AGN outburst that would be required to create these putative bubbles can be estimated using the radius of curvature of the edge {\it 4} of $\sim$130\,kpc. Combining it with the density and temperature profiles shown in \citet{ghirardini18}, the total cavity enthalpy \citep{churazov02,birzan04} is estimated at $\sim$5$\times10^{61}$\,erg. This is comparable with the most powerful cluster AGN outbursts reported in the literature \citep{mcnamara05,vantyghem14,werner16b,giacintucci20}. Such energy would be more than sufficient to completely disrupt a cool core if deposited inside the core. Note that the existence of the candidate minihalo (REF) is also in favor of the scenario that Abell~2319 used to be a cool-core cluster.

A higher-resolution radio map, sensitive to the large-scale emission and preferably obtained at low radio frequencies (where the aged AGN radio lobes should be relatively brighter), is required to determine whether this is a pair of radio lobes and not a giant halo. {\it LOFAR} may be able to derive such a map. Recently, {\it LOFAR} has resolved a presumed `giant halo' in another cluster, Abell~2390, into a pair of giant radio lobes \citep{savini19} coincident with X-ray cavities \citep{vikhlinin05}. If this scenario is supported by future radio data, it may indicate that such powerful outbursts are more common than we thought.

\subsection{A hole in the front}\label{sec:pocket}
\begin{figure}
  \centering
  \includegraphics[width=3.0in]{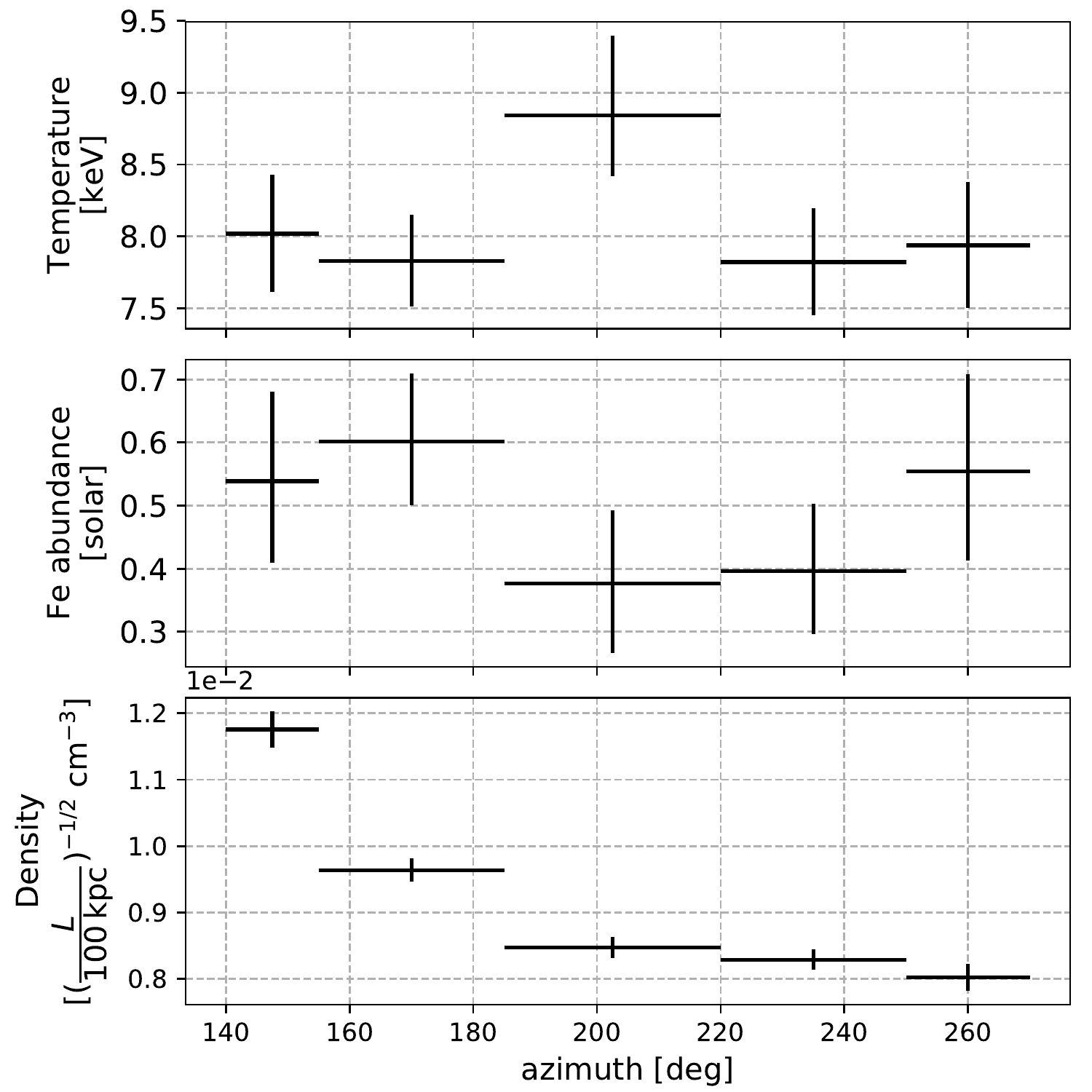}
 \caption[]{The azimuthal variations of the thermodynamic properties inside the cold front.}
 \label{img:azimuthal_variation_thermo}
\end{figure}

As we noted in Section.~\ref{sec:cf}, the amplitude of the gas density jump at the front shows a local drop around the position angle of 200$^\circ$, where the azimuthal profile of the inner power-law slope takes a local maximum (Fig.~\ref{img:azimuthal_variation}). A close examination of the X-ray image shows a curious depression in the X-ray brightness inside the cold front (dashed circle in Fig.~\ref{img:flatimage}). The gas temperature at this position is higher than elsewhere inside the cold front, see Fig.~\ref{fig:qian_tmap}, where this spot is marked by an arrow. We fit spectra in several regions inside the front shown in Fig.~\ref{img:flatimage_pocket} right (the outer radii of the sectors are determined by averaging the best-fit break radii shown in Fig.~\ref{img:azimuthal_variation}). The fit results are given in Fig.~\ref{img:azimuthal_variation_thermo}. To obtain the electron number density, we assumed the LOS depth of the X-ray emitting gas is $L=$100\,kpc. The fits confirm the higher ICM temperature at the position of the brightness depression. We offer two explanations for this feature.

\subsubsection{Non-uniform gas mixing}
In one, the higher-entropy gas in this spot (Fig.~\ref{img:thermomaps} right) originates outside the front. It could be transported under the front, for example, as a result of the KH instability \citep[e.g.,][]{roediger13b}. It is conceivable that this may happen in 3D without disrupting the sharp front boundary that we see in projection. It is expected that both the density contrast and the difference of the density slopes are smeared due to the gas mixing. The KHI-induced gas mixing is more efficient around the position angle of 200$^\circ$ than the surrounding angles, i.e. a KHI eddy may `funnel' hotter gas from the outside of the front that penetrates below the cold front interface at this location.

\subsubsection{A breach in magnetic insulation layer?}
Another interesting possibility is heat conduction from the outer side of the front. In general, the gas inside a cold front should be thermally insulated from the outer gas by a magnetic field layer parallel to the front surface, which should form as a result of magnetic draping \citep{markevitch07}. In the ICM, the heat is not transmitted across the magnetic field lines. However, MHD simulations by \citet{zuhone13a} showed that such insulation is not perfect --- some field lines can cross from the inner to the outer region. Because sloshing cold fronts have a finite `height' in the direction perpendicular to the plane of the sloshing spiral, some field lines may climb over the front `fence.' Other lines can pass through the front because of imperfect field draping. Thermal conduction along such stray magnetic field lines may heat the gas locally inside the front and produce high-entropy regions such as the one we observe.

\section{Conclusions}
In this paper, we studied the cold front in Abell~2319 in detail using $\sim$90~ks archival {\it Chandra} data. The main results of our work are summarized below.
\begin{enumerate}
\item We find several substructures associated with the cold front, including the sub-opening-angle scale variation of the interface radii of the cold front and the `finger' of cool bright gas extending outwards from the front. These features are naturally explainable by Kelvin-Helmholtz instabilities developing on the interface generated by gas sloshing mostly in the plane of the sky. Dedicated numerical simulations are required to prove the origin of the cold front and of the detected features.
\item  Thanks to the high temperature of the system, based on this interpretation, we can place an upper limit on the ICM viscosity at several times below the full isotropic Spitzer viscosity, with rather conservative assumptions about the geometry and shear velocity.
\item We found that the northern edge of the main cold front diverges into two different directions.
\item We found a concave surface brightness discontinuity southwest of the cluster core. This can result either from merger induced gas dynamics or from one of the most powerful $\sim5\times10^{61}$\,erg, AGN outbursts.
\item We found a hotter region inside the cold front. The density jump and the difference of density slopes between the gas inside and outside the interface are weaker at the corresponding azimuths. We suggest that this may be due either to non-uniform gas mixing or a hole in the magnetic layer that thermally insulates the front.
\end{enumerate}

\section*{Acknowledgements}
We thank the referee for constructive suggestions and comments, and Dr. Shutaro Ueda for useful discussions. YI is supported by the Grants-in-Aid for Scientific Research by the Japan Society for the Promotion of Science with KAKENHI grant Nos. JP18H05458, JP20K14524 and JP20K20527. A. Simionescu is supported by the Women In Science Excel (WISE) programme of the Netherlands Organisation for Scientific Research (NWO), and acknowledges the World Premier Research Center Initiative (WPI) and the Kavli IPMU for the continued hospitality. SRON Netherlands Institute for Space Research is supported financially by NWO. 

\section*{Data Availability}
The data underlying this article will be shared on reasonable request to the corresponding author.





\bibliographystyle{mnras}


\begin{thebibliography}{}
\makeatletter
\relax
\def\mn@urlcharsother{\let\do\@makeother \do\$\do\&\do\#\do\^\do\_\do\%\do\~}
\def\mn@doi{\begingroup\mn@urlcharsother \@ifnextchar [ {\mn@doi@}
  {\mn@doi@[]}}
\def\mn@doi@[#1]#2{\def\@tempa{#1}\ifx\@tempa\@empty \href
  {http://dx.doi.org/#2} {doi:#2}\else \href {http://dx.doi.org/#2} {#1}\fi
  \endgroup}
\def\mn@eprint#1#2{\mn@eprint@#1:#2::\@nil}
\def\mn@eprint@arXiv#1{\href {http://arxiv.org/abs/#1} {{\tt arXiv:#1}}}
\def\mn@eprint@dblp#1{\href {http://dblp.uni-trier.de/rec/bibtex/#1.xml}
  {dblp:#1}}
\def\mn@eprint@#1:#2:#3:#4\@nil{\def\@tempa {#1}\def\@tempb {#2}\def\@tempc
  {#3}\ifx \@tempc \@empty \let \@tempc \@tempb \let \@tempb \@tempa \fi \ifx
  \@tempb \@empty \def\@tempb {arXiv}\fi \@ifundefined
  {mn@eprint@\@tempb}{\@tempb:\@tempc}{\expandafter \expandafter \csname
  mn@eprint@\@tempb\endcsname \expandafter{\@tempc}}}

\bibitem[\protect\citeauthoryear{{Arnaud}}{{Arnaud}}{1996}]{arnaud96}
{Arnaud} K.~A.,  1996, in {Jacoby} G.~H.,  {Barnes} J.,  eds,  Astronomical
  Society of the Pacific Conference Series Vol. 101, Astronomical Data Analysis
  Software and Systems V. p.~17

\bibitem[\protect\citeauthoryear{{Ascasibar} \& {Markevitch}}{{Ascasibar} \&
  {Markevitch}}{2006}]{ascasibar06}
{Ascasibar} Y.,  {Markevitch} M.,  2006, \mn@doi [\apj] {10.1086/506508}, \href
  {http://ads.nao.ac.jp/abs/2006ApJ...650..102A} {650, 102}

\bibitem[\protect\citeauthoryear{{Birnboim}, {Keshet}  \&
  {Hernquist}}{{Birnboim} et~al.}{2010}]{birnboim10}
{Birnboim} Y.,  {Keshet} U.,   {Hernquist} L.,  2010, \mn@doi [\mnras]
  {10.1111/j.1365-2966.2010.17176.x}, \href
  {https://ui.adsabs.harvard.edu/abs/2010MNRAS.408..199B} {408, 199}

\bibitem[\protect\citeauthoryear{{B{\^\i}rzan}, {Rafferty}, {McNamara}, {Wise}
  \& {Nulsen}}{{B{\^\i}rzan} et~al.}{2004}]{birzan04}
{B{\^\i}rzan} L.,  {Rafferty} D.~A.,  {McNamara} B.~R.,  {Wise} M.~W.,
  {Nulsen} P.~E.~J.,  2004, \mn@doi [\apj] {10.1086/383519}, \href
  {https://ui.adsabs.harvard.edu/abs/2004ApJ...607..800B} {607, 800}

\bibitem[\protect\citeauthoryear{{Blanton}, {Randall}, {Clarke}, {Sarazin},
  {McNamara}, {Douglass}  \& {McDonald}}{{Blanton} et~al.}{2011}]{blanton11}
{Blanton} E.~L.,  {Randall} S.~W.,  {Clarke} T.~E.,  {Sarazin} C.~L.,
  {McNamara} B.~R.,  {Douglass} E.~M.,   {McDonald} M.,  2011, \mn@doi [\apj]
  {10.1088/0004-637X/737/2/99}, \href
  {https://ui.adsabs.harvard.edu/abs/2011ApJ...737...99B} {737, 99}

\bibitem[\protect\citeauthoryear{{Cavagnolo}, {Donahue}, {Voit}  \&
  {Sun}}{{Cavagnolo} et~al.}{2009}]{cavagnolo09}
{Cavagnolo} K.~W.,  {Donahue} M.,  {Voit} G.~M.,   {Sun} M.,  2009, \mn@doi
  [\apjs] {10.1088/0067-0049/182/1/12}, \href
  {https://ui.adsabs.harvard.edu/abs/2009ApJS..182...12C} {182, 12}

\bibitem[\protect\citeauthoryear{{Chandrasekhar}}{{Chandrasekhar}}{1961}]{chandrasekhar61}
{Chandrasekhar} S.,  1961, {Hydrodynamic and hydromagnetic stability}.
Oxford University Press

\bibitem[\protect\citeauthoryear{{Churazov}, {Sunyaev}, {Forman}  \&
  {B{\"o}hringer}}{{Churazov} et~al.}{2002}]{churazov02}
{Churazov} E.,  {Sunyaev} R.,  {Forman} W.,   {B{\"o}hringer} H.,  2002,
  \mn@doi [\mnras] {10.1046/j.1365-8711.2002.05332.x}, \href
  {https://ui.adsabs.harvard.edu/abs/2002MNRAS.332..729C} {332, 729}

\bibitem[\protect\citeauthoryear{{Churazov}, {Forman}, {Jones}  \&
  {B{\"o}hringer}}{{Churazov} et~al.}{2003}]{churazov03}
{Churazov} E.,  {Forman} W.,  {Jones} C.,   {B{\"o}hringer} H.,  2003, \mn@doi
  [\apj] {10.1086/374923}, \href {http://ads.nao.ac.jp/abs/2003ApJ...590..225C}
  {590, 225}

\bibitem[\protect\citeauthoryear{{Clarke}, {Blanton}  \& {Sarazin}}{{Clarke}
  et~al.}{2004}]{clarke04a}
{Clarke} T.~E.,  {Blanton} E.~L.,   {Sarazin} C.~L.,  2004, \mn@doi [\apj]
  {10.1086/424911}, \href
  {https://ui.adsabs.harvard.edu/abs/2004ApJ...616..178C} {616, 178}

\bibitem[\protect\citeauthoryear{{Douglass}, {Blanton}, {Randall}, {Clarke},
  {Edwards}, {Sabry}  \& {ZuHone}}{{Douglass} et~al.}{2018}]{douglass18}
{Douglass} E.~M.,  {Blanton} E.~L.,  {Randall} S.~W.,  {Clarke} T.~E.,
  {Edwards} L.~O.~V.,  {Sabry} Z.,   {ZuHone} J.~A.,  2018, \mn@doi [\apj]
  {10.3847/1538-4357/aae9e7}, \href
  {https://ui.adsabs.harvard.edu/abs/2018ApJ...868..121D} {868, 121}

\bibitem[\protect\citeauthoryear{{Eckert} et~al.,}{{Eckert}
  et~al.}{2014}]{eckert14}
{Eckert} D.,  et~al., 2014, \mn@doi [\aap] {10.1051/0004-6361/201424259}, \href
  {http://ads.nao.ac.jp/abs/2014A\%26A...570A.119E} {570, A119}

\bibitem[\protect\citeauthoryear{{Edge}, {Stewart}, {Fabian}  \&
  {Arnaud}}{{Edge} et~al.}{1990}]{edge90}
{Edge} A.~C.,  {Stewart} G.~C.,  {Fabian} A.~C.,   {Arnaud} K.~A.,  1990,
  \mnras, \href {http://ads.nao.ac.jp/abs/1990MNRAS.245..559E} {245, 559}

\bibitem[\protect\citeauthoryear{{Fabian}, {Sanders}, {Taylor}, {Allen},
  {Crawford}, {Johnstone}  \& {Iwasawa}}{{Fabian} et~al.}{2006}]{fabian06}
{Fabian} A.~C.,  {Sanders} J.~S.,  {Taylor} G.~B.,  {Allen} S.~W.,  {Crawford}
  C.~S.,  {Johnstone} R.~M.,   {Iwasawa} K.,  2006, \mn@doi [\mnras]
  {10.1111/j.1365-2966.2005.09896.x}, \href
  {http://ads.nao.ac.jp/abs/2006MNRAS.366..417F} {366, 417}

\bibitem[\protect\citeauthoryear{{Farnsworth}, {Rudnick}, {Brown}  \&
  {Brunetti}}{{Farnsworth} et~al.}{2013}]{farnsworth13}
{Farnsworth} D.,  {Rudnick} L.,  {Brown} S.,   {Brunetti} G.,  2013, \mn@doi
  [\apj] {10.1088/0004-637X/779/2/189}, \href
  {https://ui.adsabs.harvard.edu/abs/2013ApJ...779..189F} {779, 189}

\bibitem[\protect\citeauthoryear{{Feretti}, {Giovannini}  \&
  {B{\"o}hringer}}{{Feretti} et~al.}{1997}]{feretti97}
{Feretti} L.,  {Giovannini} G.,   {B{\"o}hringer} H.,  1997, \mn@doi [\na]
  {10.1016/S1384-1076(97)00034-1}, \href
  {https://ui.adsabs.harvard.edu/abs/1997NewA....2..501F} {2, 501}

\bibitem[\protect\citeauthoryear{{Foreman-Mackey}, {Hogg}, {Lang}  \&
  {Goodman}}{{Foreman-Mackey} et~al.}{2013}]{emcee}
{Foreman-Mackey} D.,  {Hogg} D.~W.,  {Lang} D.,   {Goodman} J.,  2013, \mn@doi
  [\pasp] {10.1086/670067}, \href
  {https://ui.adsabs.harvard.edu/abs/2013PASP..125..306F} {125, 306}

\bibitem[\protect\citeauthoryear{{Ghirardini}, {Ettori}, {Eckert}, {Molendi},
  {Gastaldello}, {Pointecouteau}, {Hurier}  \& {Bourdin}}{{Ghirardini}
  et~al.}{2018}]{ghirardini18}
{Ghirardini} V.,  {Ettori} S.,  {Eckert} D.,  {Molendi} S.,  {Gastaldello} F.,
  {Pointecouteau} E.,  {Hurier} G.,   {Bourdin} H.,  2018, \mn@doi [\aap]
  {10.1051/0004-6361/201731748}, \href
  {https://ui.adsabs.harvard.edu/abs/2018A&A...614A...7G} {614, A7}

\bibitem[\protect\citeauthoryear{{Ghizzardi}, {Rossetti}  \&
  {Molendi}}{{Ghizzardi} et~al.}{2010}]{ghizzardi10}
{Ghizzardi} S.,  {Rossetti} M.,   {Molendi} S.,  2010, \mn@doi [\aap]
  {10.1051/0004-6361/200912496}, \href
  {http://ads.nao.ac.jp/abs/2010A\%26A...516A..32G} {516, A32}

\bibitem[\protect\citeauthoryear{{Ghizzardi}, {De Grandi}  \&
  {Molendi}}{{Ghizzardi} et~al.}{2014}]{ghizzardi14}
{Ghizzardi} S.,  {De Grandi} S.,   {Molendi} S.,  2014, \mn@doi [\aap]
  {10.1051/0004-6361/201424016}, \href
  {https://ui.adsabs.harvard.edu/abs/2014A&A...570A.117G} {570, A117}

\bibitem[\protect\citeauthoryear{{Giacintucci}, {Markevitch}, {Cassano},
  {Venturi}, {Clarke}  \& {Brunetti}}{{Giacintucci}
  et~al.}{2017}]{giacintucci17}
{Giacintucci} S.,  {Markevitch} M.,  {Cassano} R.,  {Venturi} T.,  {Clarke}
  T.~E.,   {Brunetti} G.,  2017, \mn@doi [\apj] {10.3847/1538-4357/aa7069},
  \href {https://ui.adsabs.harvard.edu/abs/2017ApJ...841...71G} {841, 71}

\bibitem[\protect\citeauthoryear{{Giacintucci}, {Markevitch},
  {Johnston-Hollitt}, {Wik}, {Wang}  \& {Clarke}}{{Giacintucci}
  et~al.}{2020}]{giacintucci20}
{Giacintucci} S.,  {Markevitch} M.,  {Johnston-Hollitt} M.,  {Wik} D.~R.,
  {Wang} Q.~H.~S.,   {Clarke} T.~E.,  2020, \mn@doi [\apj]
  {10.3847/1538-4357/ab6a9d}, \href
  {https://ui.adsabs.harvard.edu/abs/2020ApJ...891....1G} {891, 1}

\bibitem[\protect\citeauthoryear{{Gitti}}{{Gitti}}{2015}]{gitti15}
{Gitti} M.,  2015, in Proceedings of ``The many facets of extragalactic radio
  surveys: towards new scientific challenges'' (EXTRA-RADSUR2015). 20-23
  October 2015. Bologna. p.~43

\bibitem[\protect\citeauthoryear{{Govoni}, {En{\ss}lin}, {Feretti}  \&
  {Giovannini}}{{Govoni} et~al.}{2001}]{govoni01}
{Govoni} F.,  {En{\ss}lin} T.~A.,  {Feretti} L.,   {Giovannini} G.,  2001,
  \mn@doi [\aap] {10.1051/0004-6361:20010115}, \href
  {https://ui.adsabs.harvard.edu/abs/2001A&A...369..441G} {369, 441}

\bibitem[\protect\citeauthoryear{{Govoni}, {Markevitch}, {Vikhlinin}, {van
  Speybroeck}, {Feretti}  \& {Giovannini}}{{Govoni} et~al.}{2004}]{govoni04}
{Govoni} F.,  {Markevitch} M.,  {Vikhlinin} A.,  {van Speybroeck} L.,
  {Feretti} L.,   {Giovannini} G.,  2004, \mn@doi [\apj] {10.1086/382674},
  \href {https://ui.adsabs.harvard.edu/abs/2004ApJ...605..695G} {605, 695}

\bibitem[\protect\citeauthoryear{{Harris} \& {Miley}}{{Harris} \&
  {Miley}}{1978}]{harris78}
{Harris} D.~E.,  {Miley} G.~K.,  1978, \aaps, \href
  {https://ui.adsabs.harvard.edu/abs/1978A&AS...34..117H} {34, 117}

\bibitem[\protect\citeauthoryear{{Hurier}, {Adam}  \& {Keshet}}{{Hurier}
  et~al.}{2019}]{hurier19}
{Hurier} G.,  {Adam} R.,   {Keshet} U.,  2019, \mn@doi [\aap]
  {10.1051/0004-6361/201732468}, \href
  {https://ui.adsabs.harvard.edu/abs/2019A&A...622A.136H} {622, A136}

\bibitem[\protect\citeauthoryear{Ichinohe, Werner, Simionescu, Allen, Canning,
  Ehlert, Mernier  \& Takahashi}{Ichinohe et~al.}{2015}]{ichinohe15}
Ichinohe Y.,  Werner N.,  Simionescu A.,  Allen S.~W.,  Canning R. E.~A.,
  Ehlert S.,  Mernier F.,   Takahashi T.,  2015, Monthly Notices of the Royal
  Astronomical Society, 448, 2971

\bibitem[\protect\citeauthoryear{{Ichinohe}, {Simionescu}, {Werner}  \&
  {Takahashi}}{{Ichinohe} et~al.}{2017}]{ichinohe17}
{Ichinohe} Y.,  {Simionescu} A.,  {Werner} N.,   {Takahashi} T.,  2017, \mn@doi
  [\mnras] {10.1093/mnras/stx280}, \href
  {https://ui.adsabs.harvard.edu/abs/2017MNRAS.467.3662I} {467, 3662}

\bibitem[\protect\citeauthoryear{{Ichinohe}, {Simionescu}, {Werner}, {Fabian}
  \& {Takahashi}}{{Ichinohe} et~al.}{2019}]{ichinohe19a}
{Ichinohe} Y.,  {Simionescu} A.,  {Werner} N.,  {Fabian} A.~C.,   {Takahashi}
  T.,  2019, \mn@doi [\mnras] {10.1093/mnras/sty3257}, \href
  {https://ui.adsabs.harvard.edu/abs/2019MNRAS.483.1744I} {483, 1744}

\bibitem[\protect\citeauthoryear{{Kalberla}, {Burton}, {Hartmann}, {Arnal},
  {Bajaja}, {Morras}  \& {P{\"o}ppel}}{{Kalberla} et~al.}{2005}]{kalberla05}
{Kalberla} P.~M.~W.,  {Burton} W.~B.,  {Hartmann} D.,  {Arnal} E.~M.,  {Bajaja}
  E.,  {Morras} R.,   {P{\"o}ppel} W.~G.~L.,  2005, \mn@doi [\aap]
  {10.1051/0004-6361:20041864}, \href
  {http://ads.nao.ac.jp/abs/2005A\%26A...440..775K} {440, 775}

\bibitem[\protect\citeauthoryear{{Lodders} \& {Palme}}{{Lodders} \&
  {Palme}}{2009}]{lodders09}
{Lodders} K.,  {Palme} H.,  2009, Meteoritics and Planetary Science Supplement,
  \href {https://ui.adsabs.harvard.edu/abs/2009M&PSA..72.5154L} {72, 5154}

\bibitem[\protect\citeauthoryear{{Machacek}, {Jones}, {Forman}  \&
  {Nulsen}}{{Machacek} et~al.}{2006}]{machacek06}
{Machacek} M.,  {Jones} C.,  {Forman} W.~R.,   {Nulsen} P.,  2006, \mn@doi
  [\apj] {10.1086/503350}, \href {http://ads.nao.ac.jp/abs/2006ApJ...644..155M}
  {644, 155}

\bibitem[\protect\citeauthoryear{{Markevitch}}{{Markevitch}}{1996}]{markevitch96}
{Markevitch} M.,  1996, \mn@doi [\apjl] {10.1086/310140}, \href
  {https://ui.adsabs.harvard.edu/abs/1996ApJ...465L...1M} {465, L1}

\bibitem[\protect\citeauthoryear{{Markevitch} \& {Vikhlinin}}{{Markevitch} \&
  {Vikhlinin}}{2007}]{markevitch07}
{Markevitch} M.,  {Vikhlinin} A.,  2007, \mn@doi [\physrep]
  {10.1016/j.physrep.2007.01.001}, \href
  {http://ads.nao.ac.jp/abs/2007PhR...443....1M} {443, 1}

\bibitem[\protect\citeauthoryear{{Markevitch} et~al.,}{{Markevitch}
  et~al.}{2000}]{markevitch00}
{Markevitch} M.,  et~al., 2000, \mn@doi [\apj] {10.1086/309470}, \href
  {http://ads.nao.ac.jp/abs/2000ApJ...541..542M} {541, 542}

\bibitem[\protect\citeauthoryear{{Markevitch}, {Vikhlinin}  \&
  {Mazzotta}}{{Markevitch} et~al.}{2001}]{markevitch01}
{Markevitch} M.,  {Vikhlinin} A.,   {Mazzotta} P.,  2001, \mn@doi [\apjl]
  {10.1086/337973}, \href {http://ads.nao.ac.jp/abs/2001ApJ...562L.153M} {562,
  L153}

\bibitem[\protect\citeauthoryear{{Markevitch}, {Gonzalez}, {David},
  {Vikhlinin}, {Murray}, {Forman}, {Jones}  \& {Tucker}}{{Markevitch}
  et~al.}{2002}]{markevitch02}
{Markevitch} M.,  {Gonzalez} A.~H.,  {David} L.,  {Vikhlinin} A.,  {Murray} S.,
   {Forman} W.,  {Jones} C.,   {Tucker} W.,  2002, \mn@doi [\apjl]
  {10.1086/339619}, \href {http://ads.nao.ac.jp/abs/2002ApJ...567L..27M} {567,
  L27}

\bibitem[\protect\citeauthoryear{{McNamara}, {Nulsen}, {Wise}, {Rafferty},
  {Carilli}, {Sarazin}  \& {Blanton}}{{McNamara} et~al.}{2005}]{mcnamara05}
{McNamara} B.~R.,  {Nulsen} P.~E.~J.,  {Wise} M.~W.,  {Rafferty} D.~A.,
  {Carilli} C.,  {Sarazin} C.~L.,   {Blanton} E.~L.,  2005, \mn@doi [\nat]
  {10.1038/nature03202}, \href
  {https://ui.adsabs.harvard.edu/abs/2005Natur.433...45M} {433, 45}

\bibitem[\protect\citeauthoryear{{Million} \& {Allen}}{{Million} \&
  {Allen}}{2009}]{million09}
{Million} E.~T.,  {Allen} S.~W.,  2009, \mn@doi [\mnras]
  {10.1111/j.1365-2966.2009.15359.x}, \href
  {https://ui.adsabs.harvard.edu/abs/2009MNRAS.399.1307M} {399, 1307}

\bibitem[\protect\citeauthoryear{{Nulsen}, {McNamara}, {Wise}  \&
  {David}}{{Nulsen} et~al.}{2005}]{nulsen05}
{Nulsen} P.~E.~J.,  {McNamara} B.~R.,  {Wise} M.~W.,   {David} L.~P.,  2005,
  \mn@doi [\apj] {10.1086/430845}, \href
  {https://ui.adsabs.harvard.edu/abs/2005ApJ...628..629N} {628, 629}

\bibitem[\protect\citeauthoryear{{O'Hara}, {Mohr}  \& {Guerrero}}{{O'Hara}
  et~al.}{2004}]{ohara04}
{O'Hara} T.~B.,  {Mohr} J.~J.,   {Guerrero} M.~A.,  2004, \mn@doi [\apj]
  {10.1086/382063}, \href
  {https://ui.adsabs.harvard.edu/abs/2004ApJ...604..604O} {604, 604}

\bibitem[\protect\citeauthoryear{{O'Sullivan} et~al.,}{{O'Sullivan}
  et~al.}{2012}]{osullivan12}
{O'Sullivan} E.,  et~al., 2012, \mn@doi [\mnras]
  {10.1111/j.1365-2966.2012.21459.x}, \href
  {https://ui.adsabs.harvard.edu/abs/2012MNRAS.424.2971O} {424, 2971}

\bibitem[\protect\citeauthoryear{{Oegerle}, {Hill}  \& {Fitchett}}{{Oegerle}
  et~al.}{1995}]{oegerle95}
{Oegerle} W.~R.,  {Hill} J.~M.,   {Fitchett} M.~J.,  1995, \mn@doi [\aj]
  {10.1086/117495}, \href
  {https://ui.adsabs.harvard.edu/abs/1995AJ....110...32O} {110, 32}

\bibitem[\protect\citeauthoryear{{Owers}, {Nulsen}, {Couch}  \&
  {Markevitch}}{{Owers} et~al.}{2009}]{owers09b}
{Owers} M.~S.,  {Nulsen} P.~E.~J.,  {Couch} W.~J.,   {Markevitch} M.,  2009,
  \mn@doi [\apj] {10.1088/0004-637X/704/2/1349}, \href
  {http://ads.nao.ac.jp/abs/2009ApJ...704.1349O} {704, 1349}

\bibitem[\protect\citeauthoryear{{Planck Collaboration} et~al.,}{{Planck
  Collaboration} et~al.}{2016}]{planck16xxvii}
{Planck Collaboration} et~al., 2016, \mn@doi [\aap]
  {10.1051/0004-6361/201525823}, \href
  {https://ui.adsabs.harvard.edu/abs/2016A&A...594A..27P} {594, A27}

\bibitem[\protect\citeauthoryear{{Richard-Laferri{\`e}re}
  et~al.,}{{Richard-Laferri{\`e}re} et~al.}{2020}]{richard-laferriere20}
{Richard-Laferri{\`e}re} A.,  et~al., 2020, \mn@doi [\mnras]
  {10.1093/mnras/staa2877}, \href
  {https://ui.adsabs.harvard.edu/abs/2020MNRAS.499.2934R} {499, 2934}

\bibitem[\protect\citeauthoryear{{Roediger}, {Br{\"u}ggen}, {Simionescu},
  {B{\"o}hringer}, {Churazov}  \& {Forman}}{{Roediger}
  et~al.}{2011}]{roediger11}
{Roediger} E.,  {Br{\"u}ggen} M.,  {Simionescu} A.,  {B{\"o}hringer} H.,
  {Churazov} E.,   {Forman} W.~R.,  2011, \mn@doi [\mnras]
  {10.1111/j.1365-2966.2011.18279.x}, \href
  {https://ui.adsabs.harvard.edu/abs/2011MNRAS.413.2057R} {413, 2057}

\bibitem[\protect\citeauthoryear{{Roediger}, {Kraft}, {Nulsen}, {Churazov},
  {Forman}, {Br{\"u}ggen}  \& {Kokotanekova}}{{Roediger}
  et~al.}{2013a}]{roediger13b}
{Roediger} E.,  {Kraft} R.~P.,  {Nulsen} P.,  {Churazov} E.,  {Forman} W.,
  {Br{\"u}ggen} M.,   {Kokotanekova} R.,  2013a, \mn@doi [\mnras]
  {10.1093/mnras/stt1691}, \href {http://ads.nao.ac.jp/abs/2013MNRAS.436.1721R}
  {436, 1721}

\bibitem[\protect\citeauthoryear{{Roediger}, {Kraft}, {Forman}, {Nulsen}  \&
  {Churazov}}{{Roediger} et~al.}{2013b}]{roediger13a}
{Roediger} E.,  {Kraft} R.~P.,  {Forman} W.~R.,  {Nulsen} P.~E.~J.,
  {Churazov} E.,  2013b, \mn@doi [\apj] {10.1088/0004-637X/764/1/60}, \href
  {http://ads.nao.ac.jp/abs/2013ApJ...764...60R} {764, 60}

\bibitem[\protect\citeauthoryear{{Roediger} et~al.,}{{Roediger}
  et~al.}{2015}]{roediger15b}
{Roediger} E.,  et~al., 2015, \mn@doi [\apj] {10.1088/0004-637X/806/1/104},
  \href {http://ads.nao.ac.jp/abs/2015ApJ...806..104R} {806, 104}

\bibitem[\protect\citeauthoryear{{Rossetti} \& {Molendi}}{{Rossetti} \&
  {Molendi}}{2010}]{rossetti10}
{Rossetti} M.,  {Molendi} S.,  2010, \mn@doi [\aap]
  {10.1051/0004-6361/200913156}, \href
  {https://ui.adsabs.harvard.edu/abs/2010A&A...510A..83R} {510, A83}

\bibitem[\protect\citeauthoryear{{Rossetti}, {Eckert}, {De Grandi},
  {Gastaldello}, {Ghizzardi}, {Roediger}  \& {Molendi}}{{Rossetti}
  et~al.}{2013}]{rossetti13}
{Rossetti} M.,  {Eckert} D.,  {De Grandi} S.,  {Gastaldello} F.,  {Ghizzardi}
  S.,  {Roediger} E.,   {Molendi} S.,  2013, \mn@doi [\aap]
  {10.1051/0004-6361/201321319}, \href
  {http://ads.nao.ac.jp/abs/2013A\%26A...556A..44R} {556, A44}

\bibitem[\protect\citeauthoryear{{Sanders}}{{Sanders}}{2006}]{sanders06}
{Sanders} J.~S.,  2006, \mn@doi [\mnras] {10.1111/j.1365-2966.2006.10716.x},
  \href {http://ads.nao.ac.jp/abs/2006MNRAS.371..829S} {371, 829}

\bibitem[\protect\citeauthoryear{{Sanders} et~al.,}{{Sanders}
  et~al.}{2016}]{sanders16}
{Sanders} J.~S.,  et~al., 2016, \mn@doi [\mnras] {10.1093/mnras/stv2972}, \href
  {http://ads.nao.ac.jp/abs/2016MNRAS.457...82S} {457, 82}

\bibitem[\protect\citeauthoryear{{Sarazin}}{{Sarazin}}{1986}]{sarazin86}
{Sarazin} C.~L.,  1986, \mn@doi [Reviews of Modern Physics]
  {10.1103/RevModPhys.58.1}, \href
  {http://ads.nao.ac.jp/abs/1986RvMP...58....1S} {58, 1}

\bibitem[\protect\citeauthoryear{{Savini} et~al.,}{{Savini}
  et~al.}{2019}]{savini19}
{Savini} F.,  et~al., 2019, \mn@doi [\aap] {10.1051/0004-6361/201833882}, \href
  {https://ui.adsabs.harvard.edu/abs/2019A&A...622A..24S} {622, A24}

\bibitem[\protect\citeauthoryear{{Spitzer}}{{Spitzer}}{1965}]{spitzer}
{Spitzer} L.,  1965, {Physics of fully ionized gases}.
New York: Interscience Publication

\bibitem[\protect\citeauthoryear{{Storm}, {Jeltema}  \& {Rudnick}}{{Storm}
  et~al.}{2015}]{storm15}
{Storm} E.,  {Jeltema} T.~E.,   {Rudnick} L.,  2015, \mn@doi [\mnras]
  {10.1093/mnras/stv164}, \href
  {https://ui.adsabs.harvard.edu/abs/2015MNRAS.448.2495S} {448, 2495}

\bibitem[\protect\citeauthoryear{{Su} et~al.,}{{Su} et~al.}{2017}]{su17a}
{Su} Y.,  et~al., 2017, \mn@doi [\apj] {10.3847/1538-4357/834/1/74}, \href
  {https://ui.adsabs.harvard.edu/abs/2017ApJ...834...74S} {834, 74}

\bibitem[\protect\citeauthoryear{{Sugawara}, {Takizawa}  \&
  {Nakazawa}}{{Sugawara} et~al.}{2009}]{sugawara09}
{Sugawara} C.,  {Takizawa} M.,   {Nakazawa} K.,  2009, \mn@doi [\pasj]
  {10.1093/pasj/61.6.1293}, \href
  {https://ui.adsabs.harvard.edu/abs/2009PASJ...61.1293S} {61, 1293}

\bibitem[\protect\citeauthoryear{{Sunyaev} \& {Zeldovich}}{{Sunyaev} \&
  {Zeldovich}}{1972}]{sunyaev72}
{Sunyaev} R.~A.,  {Zeldovich} Y.~B.,  1972, Comments on Astrophysics and Space
  Physics, \href {https://ui.adsabs.harvard.edu/abs/1972CoASP...4..173S} {4,
  173}

\bibitem[\protect\citeauthoryear{{Tashiro} et~al.,}{{Tashiro}
  et~al.}{2020}]{tashiro20}
{Tashiro} M.,  et~al., 2020, in Society of Photo-Optical Instrumentation
  Engineers (SPIE) Conference Series. p. 1144422, \mn@doi{10.1117/12.2565812}

\bibitem[\protect\citeauthoryear{{Ueda}, {Ichinohe}, {Kitayama}  \&
  {Umetsu}}{{Ueda} et~al.}{2019}]{ueda19}
{Ueda} S.,  {Ichinohe} Y.,  {Kitayama} T.,   {Umetsu} K.,  2019, \mn@doi [\apj]
  {10.3847/1538-4357/aafa19}, \href
  {https://ui.adsabs.harvard.edu/abs/2019ApJ...871..207U} {871, 207}

\bibitem[\protect\citeauthoryear{{Ueda}, {Ichinohe}, {Molnar}, {Umetsu}  \&
  {Kitayama}}{{Ueda} et~al.}{2020}]{ueda20}
{Ueda} S.,  {Ichinohe} Y.,  {Molnar} S.~M.,  {Umetsu} K.,   {Kitayama} T.,
  2020, \mn@doi [\apj] {10.3847/1538-4357/ab7bdc}, \href
  {https://ui.adsabs.harvard.edu/abs/2020ApJ...892..100U} {892, 100}

\bibitem[\protect\citeauthoryear{{Vantyghem}, {McNamara}, {Russell}, {Main},
  {Nulsen}, {Wise}, {Hoekstra}  \& {Gitti}}{{Vantyghem}
  et~al.}{2014}]{vantyghem14}
{Vantyghem} A.~N.,  {McNamara} B.~R.,  {Russell} H.~R.,  {Main} R.~A.,
  {Nulsen} P.~E.~J.,  {Wise} M.~W.,  {Hoekstra} H.,   {Gitti} M.,  2014,
  \mn@doi [\mnras] {10.1093/mnras/stu1030}, \href
  {https://ui.adsabs.harvard.edu/abs/2014MNRAS.442.3192V} {442, 3192}

\bibitem[\protect\citeauthoryear{{Vikhlinin}, {Markevitch}  \&
  {Murray}}{{Vikhlinin} et~al.}{2001a}]{vikhlinin01a}
{Vikhlinin} A.,  {Markevitch} M.,   {Murray} S.~S.,  2001a, \mn@doi [\apjl]
  {10.1086/319126}, \href {http://ads.nao.ac.jp/abs/2001ApJ...549L..47V} {549,
  L47}

\bibitem[\protect\citeauthoryear{{Vikhlinin}, {Markevitch}  \&
  {Murray}}{{Vikhlinin} et~al.}{2001b}]{vikhlinin01b}
{Vikhlinin} A.,  {Markevitch} M.,   {Murray} S.~S.,  2001b, \mn@doi [\apj]
  {10.1086/320078}, \href {http://ads.nao.ac.jp/abs/2001ApJ...551..160V} {551,
  160}

\bibitem[\protect\citeauthoryear{{Vikhlinin}, {Markevitch}, {Murray}, {Jones},
  {Forman}  \& {Van Speybroeck}}{{Vikhlinin} et~al.}{2005}]{vikhlinin05}
{Vikhlinin} A.,  {Markevitch} M.,  {Murray} S.~S.,  {Jones} C.,  {Forman} W.,
  {Van Speybroeck} L.,  2005, \mn@doi [\apj] {10.1086/431142}, \href
  {https://ui.adsabs.harvard.edu/abs/2005ApJ...628..655V} {628, 655}

\bibitem[\protect\citeauthoryear{{Walker}, {Sanders}  \& {Fabian}}{{Walker}
  et~al.}{2016}]{walker16}
{Walker} S.~A.,  {Sanders} J.~S.,   {Fabian} A.~C.,  2016, \mn@doi [\mnras]
  {10.1093/mnras/stw1367}, \href {http://ads.nao.ac.jp/abs/2016MNRAS.461..684W}
  {461, 684}

\bibitem[\protect\citeauthoryear{{Walker}, {Hlavacek-Larrondo},
  {Gendron-Marsolais}, {Fabian}, {Intema}, {Sanders}, {Bamford}  \& {van
  Weeren}}{{Walker} et~al.}{2017}]{walker17}
{Walker} S.~A.,  {Hlavacek-Larrondo} J.,  {Gendron-Marsolais} M.,  {Fabian}
  A.~C.,  {Intema} H.,  {Sanders} J.~S.,  {Bamford} J.~T.,   {van Weeren} R.,
  2017, \mn@doi [\mnras] {10.1093/mnras/stx640}, \href
  {https://ui.adsabs.harvard.edu/abs/2017MNRAS.468.2506W} {468, 2506}

\bibitem[\protect\citeauthoryear{{Walker}, {ZuHone}, {Fabian}  \& {Sand
  ers}}{{Walker} et~al.}{2018}]{walker18}
{Walker} S.~A.,  {ZuHone} J.,  {Fabian} A.,   {Sand ers} J.,  2018, \mn@doi
  [Nature Astronomy] {10.1038/s41550-018-0401-8}, \href
  {https://ui.adsabs.harvard.edu/abs/2018NatAs...2..292W} {2, 292}

\bibitem[\protect\citeauthoryear{{Wang}}{{Wang}}{2019}]{wangphd}
{Wang} Q. H.~S.,  2019, PhD thesis, University of Maryland

\bibitem[\protect\citeauthoryear{{Wang} \& {Markevitch}}{{Wang} \&
  {Markevitch}}{2018}]{wang18}
{Wang} Q. H.~S.,  {Markevitch} M.,  2018, \mn@doi [\apj]
  {10.3847/1538-4357/aae921}, \href
  {https://ui.adsabs.harvard.edu/abs/2018ApJ...868...45W} {868, 45}

\bibitem[\protect\citeauthoryear{{Wang}, {Markevitch}  \& {Giacintucci}}{{Wang}
  et~al.}{2016}]{wang16}
{Wang} Q. H.~S.,  {Markevitch} M.,   {Giacintucci} S.,  2016, \mn@doi [\apj]
  {10.3847/1538-4357/833/1/99}, \href
  {https://ui.adsabs.harvard.edu/abs/2016ApJ...833...99W} {833, 99}

\bibitem[\protect\citeauthoryear{{Werner} et~al.,}{{Werner}
  et~al.}{2016a}]{werner16a}
{Werner} N.,  et~al., 2016a, \mn@doi [\mnras] {10.1093/mnras/stv2358}, \href
  {http://ads.nao.ac.jp/abs/2016MNRAS.455..846W} {455, 846}

\bibitem[\protect\citeauthoryear{{Werner} et~al.,}{{Werner}
  et~al.}{2016b}]{werner16b}
{Werner} N.,  et~al., 2016b, \mn@doi [\mnras] {10.1093/mnras/stw1171}, \href
  {http://ads.nao.ac.jp/abs/2016MNRAS.460.2752W} {460, 2752}

\bibitem[\protect\citeauthoryear{{Willingale}, {Starling}, {Beardmore},
  {Tanvir}  \& {O'Brien}}{{Willingale} et~al.}{2013}]{willingale13}
{Willingale} R.,  {Starling} R.~L.~C.,  {Beardmore} A.~P.,  {Tanvir} N.~R.,
  {O'Brien} P.~T.,  2013, \mn@doi [\mnras] {10.1093/mnras/stt175}, \href
  {https://ui.adsabs.harvard.edu/abs/2013MNRAS.431..394W} {431, 394}

\bibitem[\protect\citeauthoryear{{Yan}, {Yuan}, {Zhang}  \& {Zhou}}{{Yan}
  et~al.}{2014}]{yan14}
{Yan} P.-F.,  {Yuan} Q.-R.,  {Zhang} L.,   {Zhou} X.,  2014, \mn@doi [\aj]
  {10.1088/0004-6256/147/5/106}, \href
  {https://ui.adsabs.harvard.edu/abs/2014AJ....147..106Y} {147, 106}

\bibitem[\protect\citeauthoryear{{ZuHone}, {Markevitch}  \& {Johnson}}{{ZuHone}
  et~al.}{2010}]{zuhone10}
{ZuHone} J.~A.,  {Markevitch} M.,   {Johnson} R.~E.,  2010, \mn@doi [\apj]
  {10.1088/0004-637X/717/2/908}, \href
  {https://ui.adsabs.harvard.edu/abs/2010ApJ...717..908Z} {717, 908}

\bibitem[\protect\citeauthoryear{{ZuHone}, {Markevitch}  \& {Lee}}{{ZuHone}
  et~al.}{2011}]{zuhone11}
{ZuHone} J.~A.,  {Markevitch} M.,   {Lee} D.,  2011, \mn@doi [\apj]
  {10.1088/0004-637X/743/1/16}, \href
  {http://ads.nao.ac.jp/abs/2011ApJ...743...16Z} {743, 16}

\bibitem[\protect\citeauthoryear{{ZuHone}, {Markevitch}, {Ruszkowski}  \&
  {Lee}}{{ZuHone} et~al.}{2013a}]{zuhone13a}
{ZuHone} J.~A.,  {Markevitch} M.,  {Ruszkowski} M.,   {Lee} D.,  2013a, \mn@doi
  [\apj] {10.1088/0004-637X/762/2/69}, \href
  {http://ads.nao.ac.jp/abs/2013ApJ...762...69Z} {762, 69}

\bibitem[\protect\citeauthoryear{{ZuHone}, {Markevitch}, {Brunetti}  \&
  {Giacintucci}}{{ZuHone} et~al.}{2013b}]{zuhone13b}
{ZuHone} J.~A.,  {Markevitch} M.,  {Brunetti} G.,   {Giacintucci} S.,  2013b,
  \mn@doi [\apj] {10.1088/0004-637X/762/2/78}, \href
  {https://ui.adsabs.harvard.edu/abs/2013ApJ...762...78Z} {762, 78}

\bibitem[\protect\citeauthoryear{{ZuHone}, {Kunz}, {Markevitch}, {Stone}  \&
  {Biffi}}{{ZuHone} et~al.}{2015}]{zuhone15}
{ZuHone} J.~A.,  {Kunz} M.~W.,  {Markevitch} M.,  {Stone} J.~M.,   {Biffi} V.,
  2015, \mn@doi [\apj] {10.1088/0004-637X/798/2/90}, \href
  {http://ads.nao.ac.jp/abs/2015ApJ...798...90Z} {798, 90}

\bibitem[\protect\citeauthoryear{{ZuHone}, {Miller}, {Bulbul}  \&
  {Zhuravleva}}{{ZuHone} et~al.}{2018}]{zuhone18}
{ZuHone} J.~A.,  {Miller} E.~D.,  {Bulbul} E.,   {Zhuravleva} I.,  2018,
  \mn@doi [\apj] {10.3847/1538-4357/aaa4b3}, \href
  {https://ui.adsabs.harvard.edu/abs/2018ApJ...853..180Z} {853, 180}

\bibitem[\protect\citeauthoryear{{van Weeren}, {de Gasperin}, {Akamatsu},
  {Br{\"u}ggen}, {Feretti}, {Kang}, {Stroe}  \& {Zandanel}}{{van Weeren}
  et~al.}{2019}]{vanweeren19}
{van Weeren} R.~J.,  {de Gasperin} F.,  {Akamatsu} H.,  {Br{\"u}ggen} M.,
  {Feretti} L.,  {Kang} H.,  {Stroe} A.,   {Zandanel} F.,  2019, \mn@doi [\ssr]
  {10.1007/s11214-019-0584-z}, \href
  {https://ui.adsabs.harvard.edu/abs/2019SSRv..215...16V} {215, 16}

\makeatother
\end{thebibliography}






\bsp	
\label{lastpage}
\end{document}